\documentclass[iop]{emulateapj}
\usepackage{verbatim}
\usepackage{rotating}
\usepackage{graphicx}
\usepackage{natbib}
\usepackage{todonotes}
\usepackage{color}

\begin{document}
\title{Stacking Caustic Masses from Galaxy Clusters}
\author{Daniel Gifford\altaffilmark{1}, Nicholas Kern\altaffilmark{1}, Christopher J. Miller\altaffilmark{1,2} }
\altaffiltext{1}{Department of Astronomy, University of Michigan, Ann Arbor, MI 48109 USA}
\altaffiltext{2}{Department of Physics, University of Michigan, Ann Arbor, MI 48109, USA}

\begin{abstract}

Ongoing and future spectroscopic surveys will measure numerous galaxy redshifts within tens of thousands of galaxy clusters. However, the sampling within these clusters will be low, $15 < N < 50$ per cluster. With such data, it will be difficult to achieve accurate and precise mass estimates for individual clusters using phase-space mass estimation techniques. We develop and test a new stacking algorithm based on the caustic technique, which reduces the mass scatter in $\langle \ln M_{caustic} | M_{200}\rangle$ for ensemble clusters from 70\% for individual clusters to less than 10\% for ensemble clusters with only 15 galaxies per cluster and 100 clusters per ensemble. With $> 1000$ galaxies per ensemble phase-space, the escape-velocity edge becomes readily identifiable and the presence of interloping galaxies is minimized. We develop and test an algorithm to trace the projected phase-space surface directly, which results in minimally biased dynamical mass estimates. We then quantify how binning and sampling affect the phase-space-based mass estimates when using an observational proxy that incorporates realistic mass scatter, like richness, and find the added uncertainty in the binning procedure has minimal influence on the resulting bias and scatter of the stacked mass estimates.

\end{abstract}

\section{Introduction}
\label{sec:intro}
Galaxy clusters continue to play a prominent role in our desire for precision cosmological measurements. Clusters inform us about cosmology through their abundance and spatial clustering, both of which are sensitive probes of the universe's matter density and the growth of structure due to gravity. Precision cosmological inference using clusters is only possible when we measure their characteristics as a function of mass \citep{Vikhlinin09,Rozo10}. 

Cluster mass cannot be directly observed, but its presence is visible through the gravitational potential, which can be quantified using the lensing of background galaxies (weak or strong), via the surface brightness and temperature of the intracluster medium to infer gas mass in hydrostatic equilibrium, through the maximum escape velocity surface traced by phase-spaces of the cluster galaxies, as well as through the Jean's equation \citep{Diaferio97,Meneghetti10,Carlberg97,Hoekstra15}. There are also indirect cluster mass estimation techniques, such as the X-ray gas temperature or luminosity, the SZ decrement from the scattering of the background CMB photons, the velocity dispersion, and cluster richnesses \citep{Evrard08,Andreon10,Plank11}. 

There is ongoing research to understand and control statistical and systematic uncertainties in the techniques of cluster mass inference using direct measures of the potential. Direct measurement of cluster potentials often requires a significant amount of data per cluster, whether it is the number of background galaxies for weak lensing shear profiles, X-ray photons for gas mass profiles, or spectroscopic galaxies for radius-velocity phase-space analyses. There are numerous studies which utilize simulations of idealized datasets to characterize statistical and systematic errors in direct potential measurements of individual clusters \citep{Becker11, Gruen15, Rasia06, Meneghetti10, Serra11, Geller13, Gifford13a, Gifford13b, Hoekstra15}. Another way to constrain uncertainties is to compare two different measures of the potential \cite[{\it e.g.,} ][]{Geller13, Hoekstra15}.

In reality, the quantity and quality of data quickly becomes prohibitively expensive for large samples of clusters or for faint and small clusters (e.g., low mass and/or distant). To counter the lack of data, stacking is often employed as a way to raise the signal-to-noise for an ensemble of clusters. Stacking also has the benefit of homogenizing the projected shapes in order to reduce the bias from spherically-fit profiles to non-spherical systems. 

However, stacking can just as easily induce new systematic biases. For example, \citet{Biesiadzinski12} showed how the stacked SZ signal can have significant systematic biases if the optical cluster sample selection is not correctly characterized. \citet{Becker07} used the pairwise velocity dispersion (a form of stacking) to quantify the scatter in the dispersion at fixed richness for clusters in the Sloan Digital Sky Survey data and in simulations. They recognize that the resulting stacked dispersion must be treated as non-gaussian to avoid biases in the result. Other recent studies examine how accurately stacked ensembles can infer the potential through the Jean's equation, as well as the presence of cluster shape-induced biases using both phase-space and weak lensing observables \citep{Svensmark14,Dietrich14}

Here, we study how to conduct a stacked analysis using cluster projected phase-spaces and the caustic technique to infer the escape velocity masses \citep{Diaferio97,Diaferio99}. Merging individual phase-spaces into an ensemble cluster increases the signal-to-noise of the caustic feature used to estimate the mass profile. This work extends upon \citet{Svensmark14}, who used particle data in simulations to constrain the minimum and maximum caustic mass biases caused by line-of-sight projections of the cluster phase-spaces. The \citet{Svensmark14} result is related to what was found by \citet{Dietrich14}, who used weak lensing stacking and incorporated realistic cluster selection on galaxy catalogs. In both cases, cluster shape is shown to play an important role in ensemble cluster mass estimation. What have yet to be fully investigated are the baseline accuracy of the caustic technique on stacked phase spaces, how to build ensembles clusters, and the effects of mass scatter in the binning process.

The caustic technique has been applied to stacked systems in observations before. \citet{Biviano03} stacked 43 poorly sampled clusters with galaxies out to $2r_{\text{vir}}$ and used both the caustic technique as well as a Jeans analysis to recover an average mass profile and found good agreement between the two methods. \citet{Rines03} created an ensemble cluster based on nine clusters in the CAIRNS survey. Unlike other studies, each of the included systems was sampled well enough to obtain individual measurement of velocity dispersion and $M_{200}$. They chose to scale their velocities by each system's velocity dispersion before stacking, and doing so, found agreement to within $1\sigma$ of the theoretical expectation of $M_{200} = 3\sigma^2r_{200}/G$. 

Our focus in this paper is to test the caustic technique's ability to recover the average stacked ensemble mass and the average ensemble uncertainty for different stacking strategies. We pay particular attention to the stacking algorithms, and we use realistic simulated galaxy data with projection effects that mimic what is seen in the local universe ($z \sim 0.1$). We test whether the act of stacking induces an intrinsic bias by developing a new self-stacking technique (\S \ref{sec:self_stacking}). Once the algorithm is verified, we build ensemble clusters to study how sampling rates affect the accuracy and precision of the stacked cluster masses (\S \ref{sec:massbinstack}). Finally, we use a mass-observable relation to incorporate correlated scatter into our ensemble clusters and we measure the resultant biases from mass mixing across the bin boundaries (\S \ref{sec:observablestack}). 

\section{Simulation and Semi-Analytic Catalog}
In this study, we utilize the \citet{Guo11} semi-analytic galaxy catalog and the Millennium Simulation \citep{Springel05}. Semi-analytic galaxy catalogs are built using a set of ``rules" for evolving galaxies inside identified subhalos in an N-body simulation. Subhalos are defined and located using algorithms such as SUBFIND \citep{Springel01}  which have been shown to work well to identify all of the sub-structure in N-body simulations \cite[{see also}][]{Knebe11}. The semi-analytic model captures the history of position, velocity, size, and merger history of these subhalos and applies its rule set to transform these properties into galaxy mass, circular velocity, accretion rate, and other physical parameters that can, in theory, be observed. Semi-analytic galaxy catalogs can be compared with observed luminosity functions of galaxies to judge their success.

The Millennium Database contains four different semi-analytic catalogs. One way the catalogs differ is in the way they treat ``orphan" galaxies, or semi-analytic galaxies whose host subhalo was destroyed in the simulation. Orphan galaxy treatment is necessary since subhalos are too easily destroyed in N-body simulations due to limited resolution. To combat this, once a subhalo is destroyed, the semi-analytic galaxy lives on for a time following the most bound particle of the destroyed subhalo. Dynamical friction arguments are applied to determine the time the orphan galaxy exists in the simulation. \citet{Gifford13b} studied these semi-analytic catalogs and found the Guo catalog galaxy cluster velocity dispersion and mass estimates to be near the average of the catalogs and unbiased using a set of assumptions for mass estimation.

\begin{figure*}
\plottwo{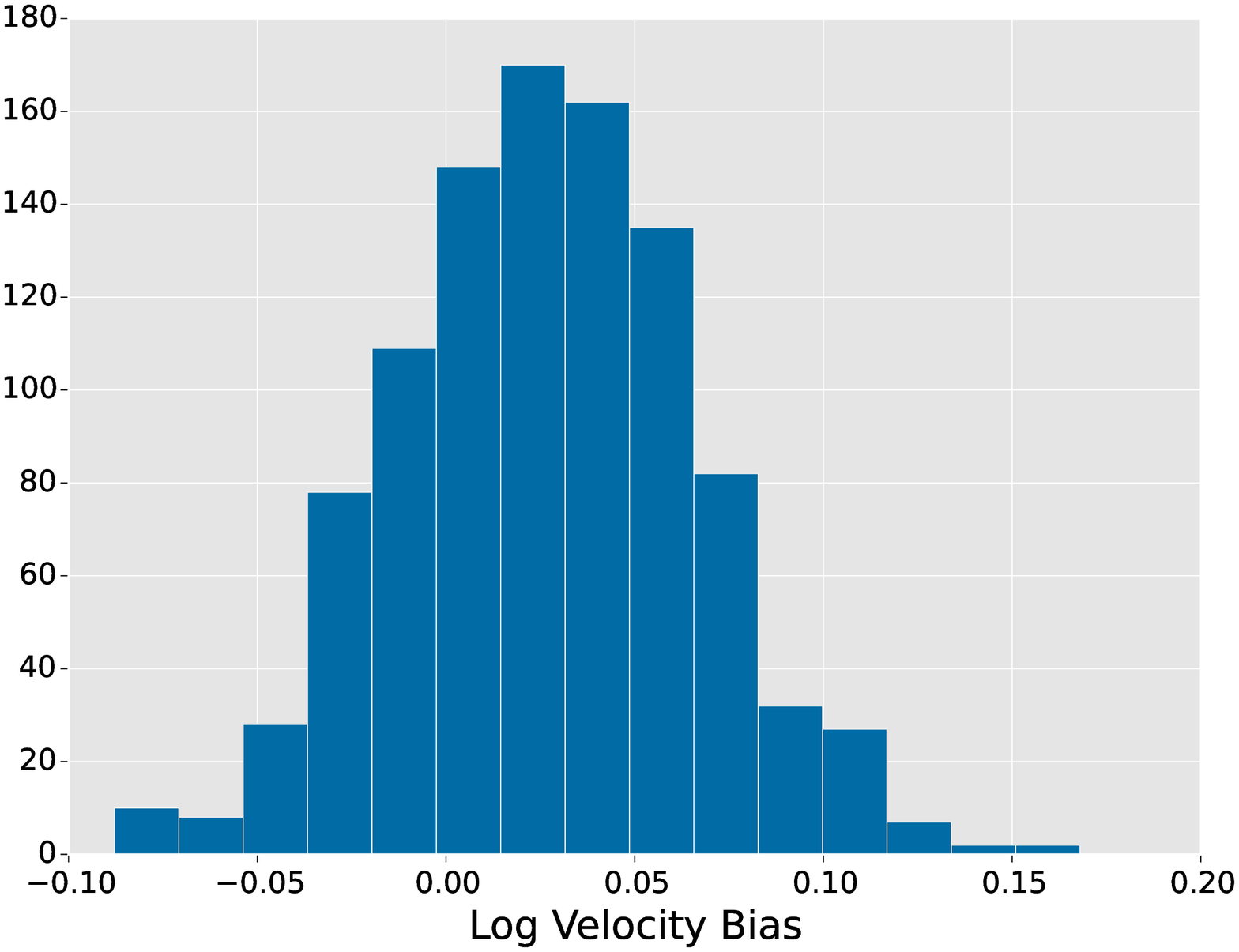}{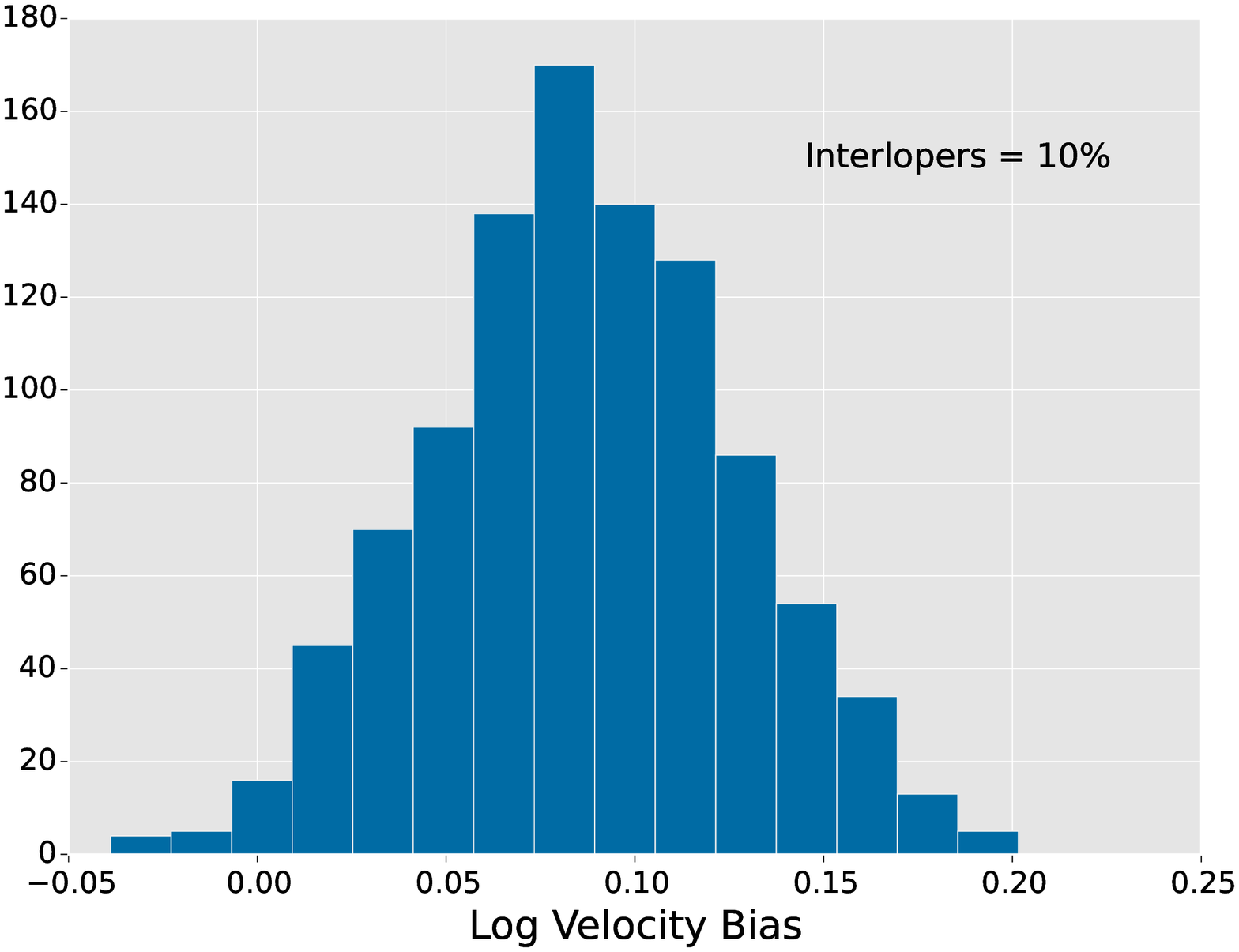}
\caption{The expected log bias in velocity dispersion simulating 15 draws from 50 gaussians which themselves are randomly sampled from a log-normal distribution with a scatter of 20\%. (Left) The resulting velocity bias compared with the median dispersion of the 50 input gaussians is $\sim 3-5\%$. (Right) If we assume 10\% of the galaxies are instead drawn from a uniform distribution representing interlopers, the bias increases to $\sim 10\%$. \label{fig:mixture_bias}}
\end{figure*}

In sections \ref{sec:self_stacking} and \ref{sec:self_stacking_r} our sample contains 20 clusters from the Guo semianalytic catalog at redshift = 0 for which we have 3-dimensional position and velocity information. These clusters are located within the central $50 \times 50 \times 50 h^{-1}$ Mpc of the $500 h^{-1}$ Mpc per side simulation box. This allows for long lines-of-sight to each system of at least $200 h^{-1}$ Mpc. Long lines-of-sight are necessary to build phase spaces with similar projected interloper properties of clusters at low redshift.

The remaining sections utilize the Henriques all-sky light cone \citep{Henriques12,Guo11} to study stacking independent clusters. This sample consists of $6000$ clusters that span a mass range of $7.2 \times 10^{13} - 2.1 \times 10^{15} h^{-1} M_{\odot}$. The redshift range is selected to be between $0.0 - 0.15$ to mimic shallow-wide surveys. This large sample allows us to test for systematics over a large cluster mass range. In all our analyses, we assume a flat $\Lambda$CDM cosmology with $\Omega_{\text{M}} = 0.25$, $\Omega_{\Lambda} = 0.75$, and $H = 100 h$ km s$^{-1}$ Mpc$^{-1}$.

\section{The Caustic Technique} \label{sec:caustic_tech}
Under Newtonian dynamics, the escape velocity of a spherically symmetric cluster relates to its gravitational potential by
\begin{equation}
v^2_{esc}(R) = -2\Phi(R),
\label{eq:escape_newton}
\end{equation}
where $R$ is the 3-dimensional measured distance to the center of the cluster. If the dynamics of the system are controlled by the gravitational potential, galaxies which cannot escape the potential exist in a well-defined region of radius/velocity $(R-v)$ phase space. The extrema of the velocities in this phase space define an edge, the escape velocity profile $v_{esc}(R)$, which can also be observed directly with 3 dimensional position and velocity information, or estimated using projected sky coordinates and line-of-sight velocities as $v_{esc}(r)$. Given the observed $v_{esc}(r)$, the escape velocity or ``caustic" technique allows one to infer the mass profile of a cluster to well beyond the virial radius \citep{Diaferio97}. Once the escape velocity surface is identified, the equation
\begin{equation}
GM(<r_{200}) = \mathcal{F}_{\beta} \int^{r_{200}}_{0} v_{esc}^2(r') dr'
\label{eq:caustic_eq}
\end{equation}
calculates the mass within the radius $r_{200}$ where $v_{esc}^2(r)$ is the observed (projected) escape velocity profile and $\mathcal{F}_{\beta}$ is a function which depends on the density, the potential, and the projected anisotropy profile. It is usually approximated as a constant and calibrated through simulations. 

\subsection{Calibrating the Caustic Masses}
\label{sec:calib1}
\citet{Gifford13a} quantify how well the average calibration constant, $\mathcal{F}_{\beta}$, can be determined for individual systems. These calibrations involve choosing the correct ``surface'' to prescribe as tracing the projected escape velocity and in defining a value for $\mathcal{F}_{\beta}$. As discussed in detail in \citet{Diaferio99},
the constant $\mathcal{F}_{\beta}$ is derived as a function of the dark matter density ($\rho(r)$) and potential ($\Phi(r)$) profiles, as well as the galaxies' velocity anisotropy profile $\beta (r) = 1 - (v_{\theta}^2 + v_{\phi}^2) / v_{r}^{2}$. Analytically, $\mathcal{F}_{\beta} (r)$ is expressed as:
\begin{equation} \label{eqn:Fbeta}
\mathcal{F}_{\beta} (r) = \frac{-2\pi G \rho (r) r^2}{\Phi (r)} \frac{3-2\beta (r)}{1-\beta (r)}.
\label{eq:avgfb}
\end{equation}
Considering density profiles of halos and clusters in simulations can be modeled in the functional form defined by \citet{NFW97}, we can reduce equation \ref{eqn:Fbeta} to
\begin{equation} \label{eq:avgfb_nfw}
\mathcal{F}_{\beta,NFW}(r) = \frac{c^{2} s^2}{2 \ln (1 + cs) (1+cs)^2} \frac{3-2\beta (r)}{1-\beta (r)},
\end{equation}
where $s = r / r_{200}$ and $c = r_{200} / r_s$ such that $r_s$ is the NFW scale radius. The NFW profile is not the only parametrization of the density found in literature. For example, The ``Einasto Profile" \citep{Einasto69} has been shown to be a better fit to the density profiles outside the virial radius of clusters \citep{Miller16}. 
\citet{Dehnen93} uses a different functional (a generalized Jaffe profile) which, like the Einasto profile, models a steeper density profile outside the virial radius as compared with the NFW functional form. Due to close similarities between the generalized Jaffe profiles and the Einasto profiles, we focus on results from the NFW and Einasto parametrizations.

\citet{Diaferio99} and \citet{Serra11} approximate the radial average of equation \ref{eq:avgfb} as a constant ($\mathcal{F}_{\beta}$) out to 2-3$\times$r$_{200}$, noting that its value varies slowly with radius. When defined as a constant over some range in radius, 
$\mathcal{F}_{\beta}$ acts as a caustic mass calibrator, which depends weakly on variations in concentration and the radially averaged velocity anisotropy $\beta$.

There is some debate in the literature about the value of $\mathcal{F}_{\beta}$, with values ranging from 0.5 - 0.7 \citep{Diaferio99, Serra11, Gifford13b, Svensmark14}. Nominally, this term resides within the integral in Equation \ref{eq:caustic_eq} and so the correct value is determined through its nearly constant radial average. As shown in \citet{Gifford13a} and \citet{Diaferio99}, large variations in concentration and $\beta$ lead to small variations in the $\mathcal{F}_{\beta}$ and caustic masses, so these are not the source of the wide range in accepted $\mathcal{F}_{\beta}$ values. Instead, \citet{Gifford13b} show that the differences can mostly be attributed to the use of different dynamical tracers (i.e., velocity bias), which enters into the technique when the surface is determined, i.e., equation \ref{eq:caustic_eq}, which is derived by requiring equation \ref{eq:escape_newton}.

All caustic analyses on clusters utilize a surface calibration based on
\citet{Binney87}, who show that 
\begin{equation} \label{eqn:Binney}
\langle v_{esc}^2 \rangle = \frac{-4 W}{M}
\end{equation} 
where $M$ and $W$ are the total mass and potential energy of the system respectively. If the system is in virial equilibrium, $-W = 2K$, where the total kinetic energy $K = 1/2 M \langle v^2 \rangle$, then we can identify the escape velocity profile by choosing the iso-density contour in phase-space that satisfies the equation
\begin{equation}
\label{eqn:surface}
\langle v_{esc}^2 \rangle - 4 \langle v^2 \rangle = 0.
\end{equation}
If $\langle v^2 \rangle$ is calculated using biased dynamical tracers, the inferred $v_{esc}(r)$ will also be biased. This can be compensated for by calibrating $\mathcal{F}_{\beta}$ to recover unbiased masses.

It is not necessary that $\mathcal{F}_{\beta}$ be treated as a free-parameter to calibrate unbiased cluster masses against simulations \citep{Alpaslan12,Gifford13b,Svensmark14}. Utilizing density-potential pair equations for either the NFW or Einasto profiles \citep{Miller16} and assuming some value for the radially averaged anisotropy parameter $\beta$, one can calculate $\mathcal{F}_{\beta}$ from equation \ref{eq:avgfb} or \ref{eq:avgfb_nfw} directly, so long as one is given a cluster mass. We use $M_{200}$ and $r_{200}$ from our simulated cluster sample along with the concentration-mass relation from \citet{Merten15} to analytically model the NFW density and potential profiles given each cluster's mass and inferred concentration. The Einasto density profiles are derived by fitting to the analytic NFW density profiles within $r_{200}$ \citep{Sereno16, Stark16}. Finally, we model the anisotropy as a constant inside $r_{200}$ with $\beta = 0.15 \pm 0.10$ \citep{Iannuzzi12}. Marginalizing over all uncertainties in $c$ and $\beta$ results in a suite of $\mathcal{F}_{\beta} (r)$ profiles for each parametrization that can be radially averaged into the constant $\mathcal{F}_{\beta} = \int_0^r \mathcal{F}_{\beta}(x)dx / r$ per cluster. We use all of the clusters to find a median value of $\mathcal{F}_{\beta} = 0.56 \pm 0.05$ within $r_{200}$ assuming an NFW profile and $\mathcal{F}_{\beta} = 0.63 \pm 0.05$ when using the Einasto profile. We emphasize that the values for $\mathcal{F}_{\beta}$ are not from fits to any data, but are the result of models given the list of cluster $M_{200}$s and $r_{200}$s. 

\citet{Serra11} utilize numerically evaluated potentials, which they show are $\sim 10\%$ lower than the potentials expected from the NFW density profile via the Poisson equation. The challenge in this approach is to carefully match the potential in equations \ref{eq:avgfb} or \ref{eq:avgfb_nfw} to the escape surface. This is because equation \ref{eqn:surface} requires that the density and potential profiles be Poisson pairs \citep{Binney87}. The $\sim 10\%$ difference between our value of $\mathcal{F}_{\beta}$ = 0.63 and the value in \citet{Serra11} of $\mathcal{F}_{\beta}$ = 0.7 is explained through our use of a Poisson potential (with potential relative to infinity) and \citet{Serra11} use of a numerical potential (with potential relative to 10Mpc), given that the average densities are the same. We have explored other possibilities such as radial and mass dependencies of $\mathcal{F}_{\beta}$ and find they account for differences of a few percent. 

Taking all of these issues into account, we conclude that the literature values of $\mathcal{F}_{\beta}$ are quite consistent ($0.6 \le \mathcal{F}_{\beta} \le 0.65$) between techniques which utilize equations \ref{eq:avgfb} or \ref{eq:avgfb_nfw} \cite[{\it e.g.,} ][]{Diaferio99,Serra11}, and techniques which calibrate it to individual systems using $\Lambda$CDM simulations \cite[{\it e.g.,} ][]{Gifford13b,Svensmark14}.

Another challenge when using stacked ensemble clusters and the caustic technique is in the surface calibration (e.g., equation \ref{eqn:surface}.) As shown in \citet{Becker07} and as we highlight below, the velocity dispersion of a stacked phase-space is a biased representation of the mean of the underlying velocity dispersion. Therefore in \S \ref{sec:esc_method}, we develop a new technique to calibrate the escape-velocity surface which leverages the high sampling of the stacked phase-spaces. 

\subsection{Estimating the Caustic Profile} \label{sec:esc_method}
The caustic technique is an estimate of the projected escape velocity profile such that the caustic profile $v_{esc} (r)$ and the potential profile $\Phi (r)$ (in the definition of $\mathcal{F}_{\beta}$) must be equal. As the name suggests, the caustic should be a sharp density drop-off in the cluster projected phase-space; however, with a sampling of even 100 galaxies inside the virial radius it is difficult to identify this edge by eye, let alone algorithmically. 

This challenge has been overcome by identifying cluster members, measuring the cluster velocity dispersion, and using a virialization condition to calibrate the caustic surface based on a series of measured iso-density contours \citep{Diaferio99}. This step makes caustic mass estimates of individual clusters possible without the need to visually identify a sharp edge in projection. The downside of using equation \ref{eqn:surface} is that we add an additional assumption to our methodology: that the stacked line-of-sight velocity dispersion represents an average of the individual dispersions. In velocity space, stacking is analogous to mixing semi-randomly sampled gaussians if each cluster's velocity dispersion is approximately normal. On average, the resulting stacked distribution will have a velocity dispersion that is larger than the mean of the individual dispersions, or $\sigma_{\text{stack}} > \langle \sigma_i \rangle$. We show this by simulating draws from mixed gaussian distributions and reporting the log difference between the dispersion of the sample and the average dispersion of the input gaussian population. 

In Figure \ref{fig:mixture_bias}, we simulate the expected bias for typical observational values of 20\% scatter in $\sigma_i$, 15 random samples per Gaussian, and 50 Gaussians per stack in 1000 experiments. The result is an expected velocity dispersion bias of 3-5\% (left figure). This does not include a potential background interloper population uniformly distributed in velocity space that would work to increase this bias. \citet{Biviano06} find an interloper fraction of 18-25\% when using sigma-clipping techniques on randomly chosen dark matter particles; however, we expect our values to be lower given our more complex shifting-gapper approach which also utilizes magnitude information \citep{Gifford13b}.

We re-quantify the velocity dispersion bias when 10\% of the galaxies are drawn from a uniform background population in velocity space. We find that the bias increases to $\sim 10\%$. The translation from velocity dispersion bias to mass bias in the caustic technique is complicated and discussed further in \S \ref{sec:massbinstack}.

Given the expected velocity bias in our ensemble clusters and our goal to remove the virial condition from the technique, we wish to eliminate the use of equation \ref{eqn:surface} altogether. We can do this through a direct identification of the caustic surface in highly sampled phase-spaces. Stacking allows us to overcome the lack of signal in the cluster projected phase-spaces and identify a caustic edge, despite projection blurring effects. 

Using the radius/velocity phase-space, we identify iso-density surfaces based on medians of velocity percentiles in the radially binned velocity distributions using a mirrored phase-space \cite[{see} ][]{Gifford13b}. Interloper rejection is done in two phases. First, a shifting-gapper algorithm is run to eliminate obvious outliers in the phase-spaces \cite[{see}][]{Gifford13b}. However, because our phase-spaces are densely populated, we also make cuts in velocity based on smoothed phase-space density. In each radial bin, we make a cut in absolute velocity where the smoothed density reaches the estimated background density which differs from stack-to-stack. This usually occurs between 2-3$\times$ the velocity dispersion in each radial bin. 

Once we have removed potential interlopers, instead of using equation \ref{eqn:surface} we use the analytically calculated Einasto potential profiles to calibrate the correct iso-density contour selected by velocity percentile. We find that the iso-density contour matched to the median of galaxies with velocities above the $\sim$90th percentile (within radial bins $< \langle r_{200} \rangle$ of the ensemble) recovers the mean Einasto potential profile for each ensemble. This new surface calibration technique is independent of mass and sampling, at least for the systems in our data set. Most importantly, this algorithm ensures that our caustic surface matches the analytic potential, as required in equation \ref{eq:caustic_eq}.

As a result of projection effects, the NFW surface calibration is very similar to the Einasto profile and is within one percentile of the 90th percentile chosen above. We note that while the NFW escape velocity profile is slightly higher than the Einasto profile for a given stacked cluster, the smaller $\mathcal{F}_{\beta,NFW}$ as compared with $\mathcal{F}_{\beta,Einasto}$ balance out in equation \ref{eq:caustic_eq} which creates nearly identical results of bias and scatter as presented in \S \ref{sec:Stacking} when using the Einasto profile.  We note that Miller et al. (2016) showed that the Einasto and the generalized Jaffe density-potential pairs perform nearly identically well, and so we can surmise that this calibration holds for at least three popular potential-density Poisson pairs.

Unlike Serra et al. (2011) who utilize
numerical potentials determined from particle information on individual clusters, we use an analytical representation of the potential and pay strict attention to the \citet{Binney87}
framework to infer caustic surfaces in stacked phase-spaces.
Since our analysis is based upon a semi-analytic representation of the galaxies
in a light-cone simulation, we cannot numerically evaluate the projected particle-based
potential-density pairs. 

Our emphasis on the Einasto profile has several advantages.
First, it ensures a true density-potential pairing \citep{Binney87}. Second, the Einasto, like the NFW, is
a well-known representation of the average (or stacked) density
profile of collapsed halos to $r_{200}$ \citep{NFW97}. Third, the Einasto 
profile may actually provide better fits to the density and potential profiles both inside and outside 
the virial radius of galaxy cluster size halos \citep{Merritt06,Miller16}. Fourth, the Einasto profile has been tested observationally (Sereno et al. 2016, Stark et al. 2016). However, we caution that 
other potential-density pair representations of the cluster profiles may do a better job than
what we find for the Einasto.

At the same time, our emphasis on the Einasto (or any) analytical representation
has its own pitfalls. While the Einasto profile is a good fit to simulated and
real mass density profiles of clusters, this obviously may not translate to observed
dark matter halos. Likewise, if a cluster sample is dominated by
non-virialized systems, parametrizations like the Einasto and the NFW are certainly not appropriate, since they
were both constructed against collapsed (or virialized) systems.

\begin{figure}
\plotone{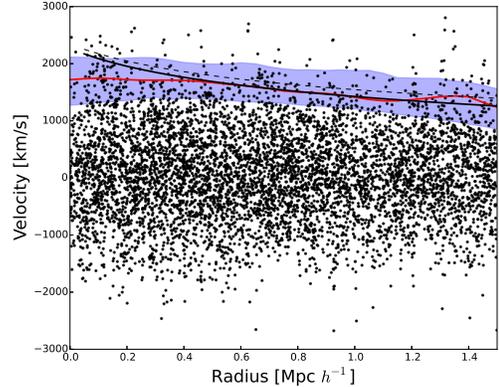}
\caption{A stacked cluster projected phase-space of the top 50 brightest galaxies from the most massive 100 clusters in the sample. The solid black line is the analytical $\sqrt{-2\Phi_{Einasto}}$ and the thin dashed line is the analytical NFW equivalent using the average properties of the stack. The red line is the inferred caustic surface determined by matching the iso-density contour to the median of the radially binned line-of-sight velocities for those above the 90th percentile of galaxy velocities over 6 radial bins. The blue band contains 68\% of the caustic surfaces estimated by each individual cluster in the stack. \label{fig:stacked_phase}}
\end{figure}

In Figure \ref{fig:stacked_phase}, we show a stacked cluster projected phase-space from our simulations. The stack is built by sampling the top 50 brightest galaxies from the most massive 100 clusters in the sample. The red line shows the caustic edge selected using line-of-sight velocities above the 90th percentile of observed velocities across 6 radial bins within the $\langle {r_{200}} \rangle$ of this ensemble. This is compared with the analytically calculated $\sqrt{-2\Phi_{Einasto} (r)}$ (solid black line) and the NFW equivalent (dashed black line) using the average properties of the system.  The difference between the model NFW and Einasto escape velocity edge is small, and switching between the two models changes the inferred precentile by less than a percent. However, the Einasto result matches what is expected by a direct calibration of $\mathcal{F}_{\beta}$ \citep{Gifford13b, Svensmark14}. The blue band encompasses the 68\% distribution of the individual caustic profiles estimated for each input cluster to the stack using the velocity dispersion to calibrate each caustic surface. Because these massive clusters have high sampling, the individual surfaces agree with the stacked surface within the scatter. 
\begin{figure*}
\plottwo{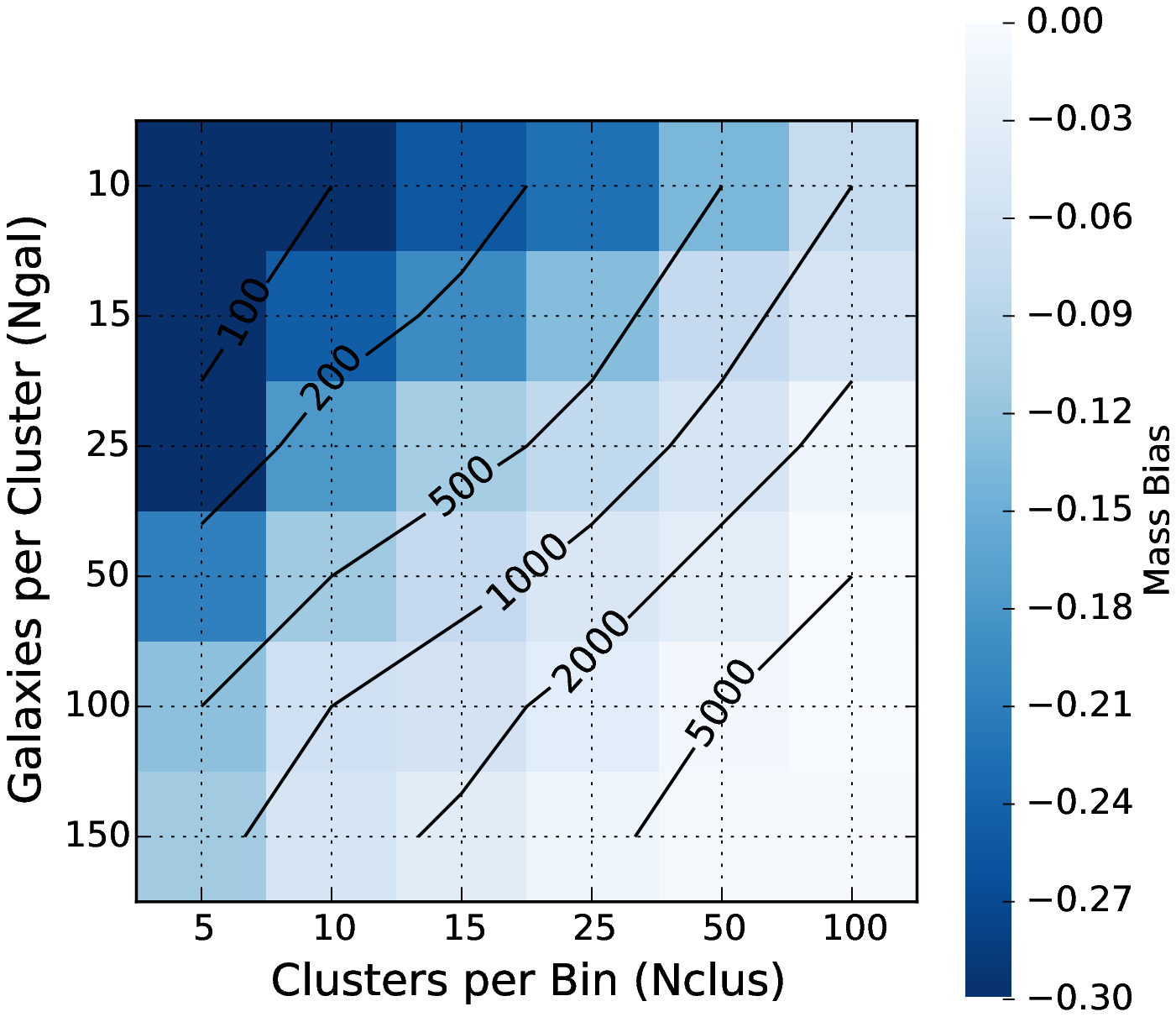}{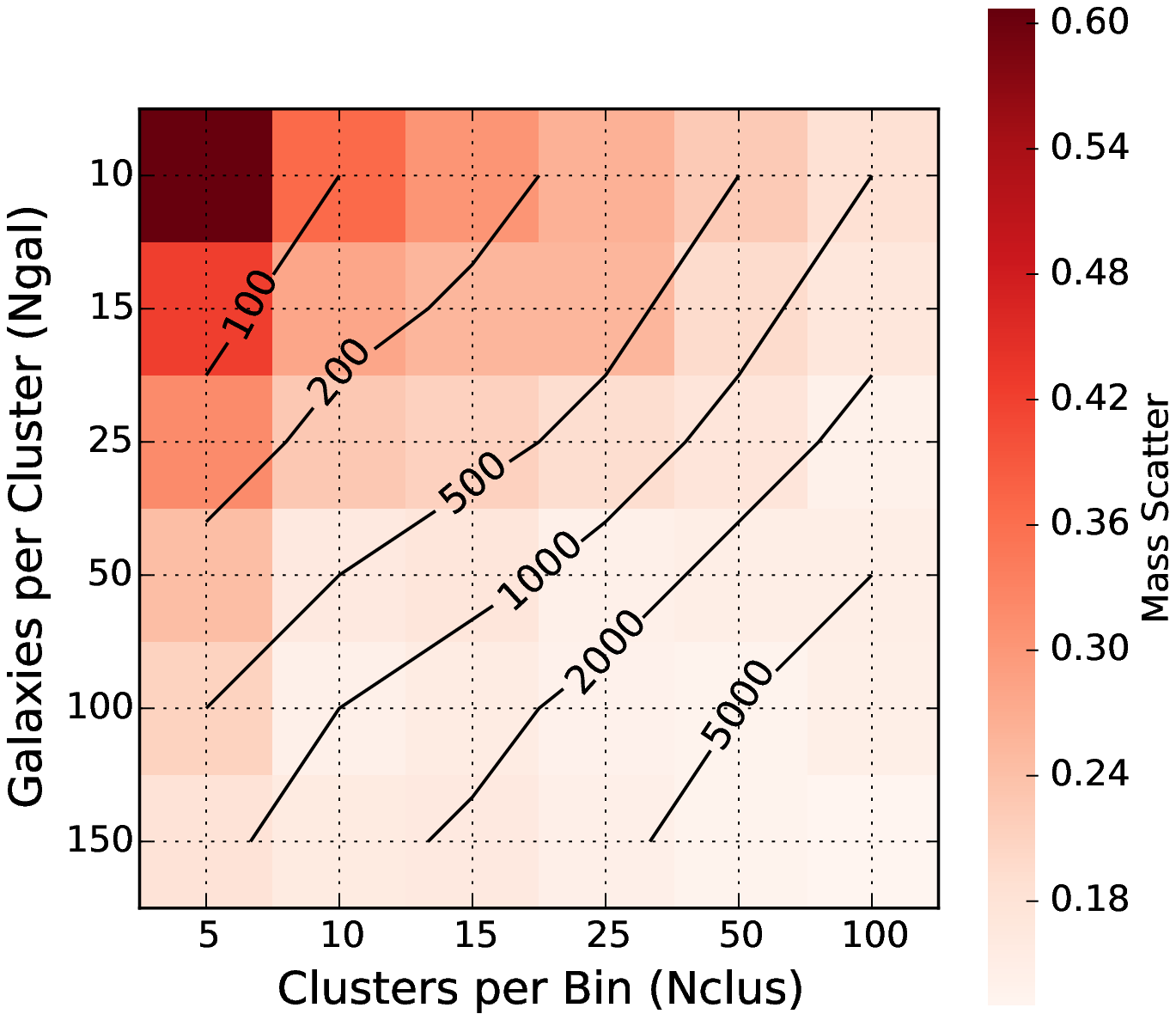}
\caption{Left: We show caustic mass bias for self-stacked ensembles as a function of the number of random lines-of-sight to each cluster and the number of galaxies sampled from each line-of-sight. The labeled diagonal lines are contours of constant stacked phase-space richness $N_{ens}$. Right: The caustic mass scatter for self-stacked ensembles as a function of the number of random lines-of-sight to each cluster and the number of galaxies sampled from each line-of-sight. \label{fig:SS_Bias_Scatter}}
\end{figure*}

In \S \ref{sec:Stacking}, we test the ability of stacking to recover the average cluster mass in stacked ensemble phase-spaces by adjusting the number of galaxies per cluster and the number of clusters used in stacking. We will also study how the cluster binning procedure affects the stacked mass estimates. 

\section{Stacking Methods and Results} \label{sec:Stacking}
In sub-sections \ref{sec:self_stacking}, \ref{sec:massbinstack}, and \ref{sec:observablestack} we investigate and quantify the overall systematic uncertainties when using stacking cluster projected phase-spaces and our revised caustic technique to estimate average masses. In \S \ref{sec:self_stacking} we introduce a way to stack single clusters in order to test the fundamental basis of our algorithm. We then introduce mass-mixing to our stacking procedure in \S \ref{sec:massbinstack} by creating ensembles of clusters through binning directly on mass. Finally in \S \ref{sec:observablestack}, we build cluster ensembles using a mass proxy: the projected richness estimates of each cluster which has realistic scatter.

\subsection{Self-stacking} \label{sec:self_stacking}
Stacking cluster phase-spaces first requires a decision on how to ``bin" or stack on cluster properties. Ideally, we first stack clusters of identical mass (and concentration) in order to eliminate the overall mass-mixing in each bin. This allows us to test for any potential systematic biases in the technique itself. However, this task is difficult. Ideally, one needs multiple realizations (or re-simulations) of the same cosmology in order to achieve enough statistics without binning. 

To solve this problem for our sample of clusters based in the Millennium Simulation, we devise a technique that stacks cluster projected phase-spaces with mass bins of infinitesimal width. This is achieved by stacking multiple lines-of-sight to one cluster in order to build a stacked phase-space. We term this technique self-stacking. The inferred stacked masses are then compared to a single well-defined mass to identify biases in the technique.

\subsubsection{Self-stacking Methods} \label{sec:self_stacking_r}
Technically, self-stacking mimics stacking different individual systems. We treat each line-of-sight projection to a single cluster as a unique observation which, when stacked with $N_{los}$ random projections, produces an ensemble phase-space that we use to identify the caustic profile. The exact steps are as follows:
\begin{enumerate}
\item $N_{los}$ random lines-of-sight to a cluster are chosen, the galaxies are projected to create $N_{los}$ radius-velocity projected phase-spaces.
\item $N_{gal}$ galaxies are chosen randomly from the top $N_{bright}$ brightest galaxies projected within the virial radius of the cluster in each phase-space. The $N_{gal}$ brightest galaxies are chosen randomly to avoid artificial structure in the projected phase-space due to using the same cluster.
\item The stacked phase-space will contain the combined projected positions and velocities of $N_{ens} = N_{los} \times N_{gal}$ galaxies.
\item The caustic profile is identified and mass estimated for each stacked system.
\end{enumerate}

How we choose $N_{gal}$ from the top $N_{bright}$ galaxies is carefully considered. There are several ways to achieve a desired sampling. First, along each line-of-sight to a cluster, galaxies can be rank ordered by a chosen magnitude (e.g. SDSS r-band) and the brightest galaxies are selected only from the top of the list ($N_{gal} = N_{bright}$). This method's advantage lies in its closeness to realistic spectroscopic follow-up. Usually, the brightest galaxies in any given field are given preference in spectroscopic surveys for practical purposes in observation and reduction. 

The problem with sampling the same bright galaxies during the self-stacking process in simulations is the repeated measurements along different lines-of-sight. Due to the simple projection geometry, a single galaxy's projected distance from the cluster center changes very slowly as the observer's line-of-sight to the cluster shifts. If the observer plots the cluster projected phase-space position of a single galaxy for many different random lines-of-sight, the phase space will show strong vertical structure bounded by a maximum in projected distance equal to the 3D distance from the galaxy to the cluster center. When sampling many galaxies per line-of-sight, this artificial structure can heavily influence the phase space density, and consequently, the iso-density contour that defines the caustic profile. Ultimately, this leads to non-physical phase-spaces.

Another method of selecting galaxies along different lines-of-sight also requires sorting the galaxies brighter than some magnitude limit to create a list $N_{bright}$. However, instead of always taking the $N_{gal}$ brightest galaxies from each line-of-sight, a fraction of the sorted list is selected in a random fashion. For example, if we wish to select 10 galaxies along each line-of-sight to a cluster to be stacked into the final ensemble, we could randomly choose these galaxies from a list of the 100 brightest galaxies. This does not ensure complete uniqueness of galaxies in the final ensemble, but the frequency of repeated draws will become very low as $N_{gal} \ll N_{bright}$. Choosing the fractional difference between $N_{gal}$ and $N_{bright}$ using this method requires us to strike a balance. If we impose that the fraction be large ($N_{gal} \approx N_{bright}$), then there will be more artificial structure due to the repeated selection of galaxies like in the previous method. If instead we impose that the fraction be small ($N_{gal} \ll N_{bright}$), then galaxies will be randomly selected from a very large and therefore increasingly faint set of galaxies that may not be realistic for typical observed clusters. This can also increase the interloper-to-member ratio of galaxies in the phase-space which works to further blur the caustic edge in projection and increase the uncertainty in its position.

\begin{figure*}
\plottwo{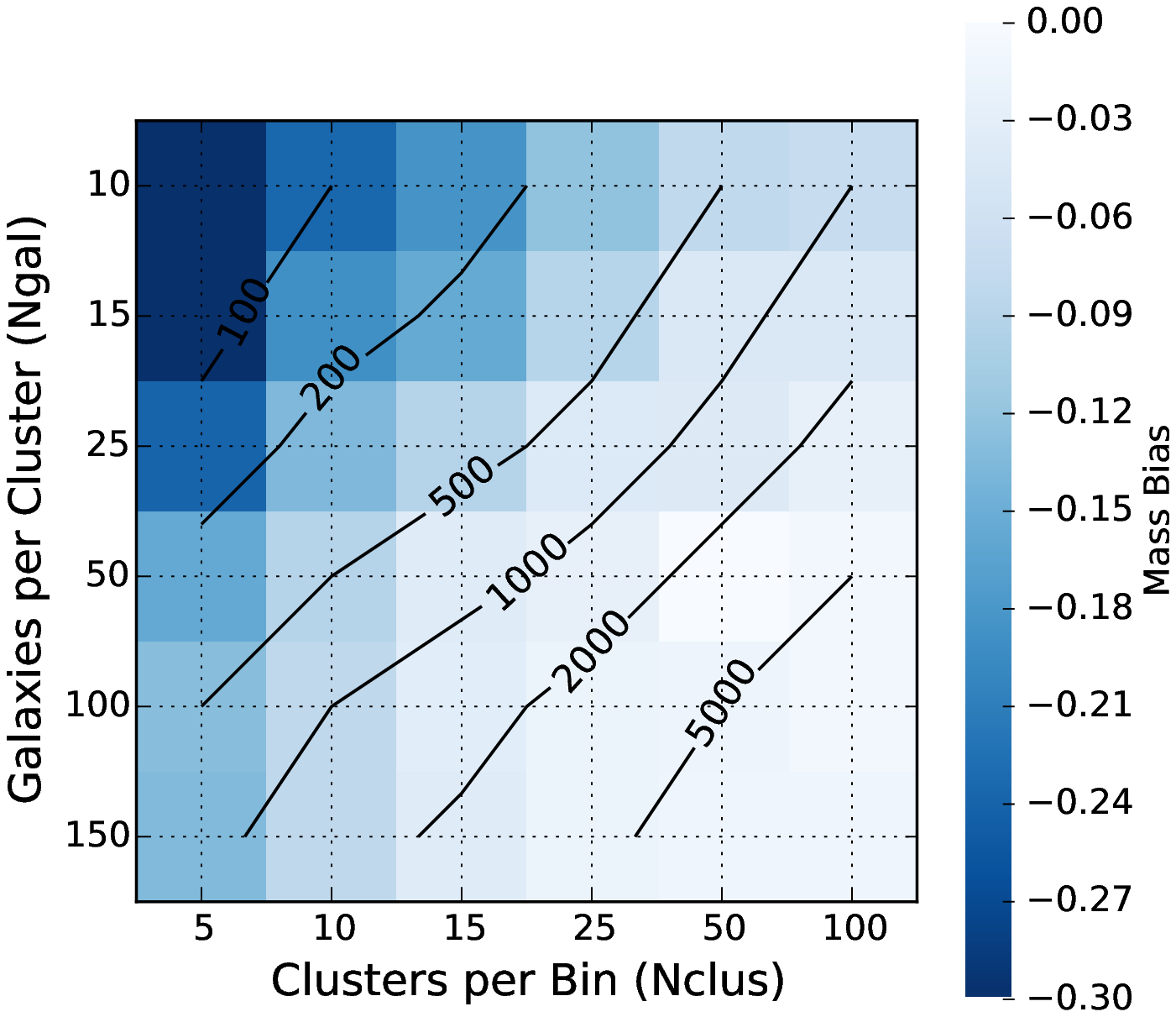}{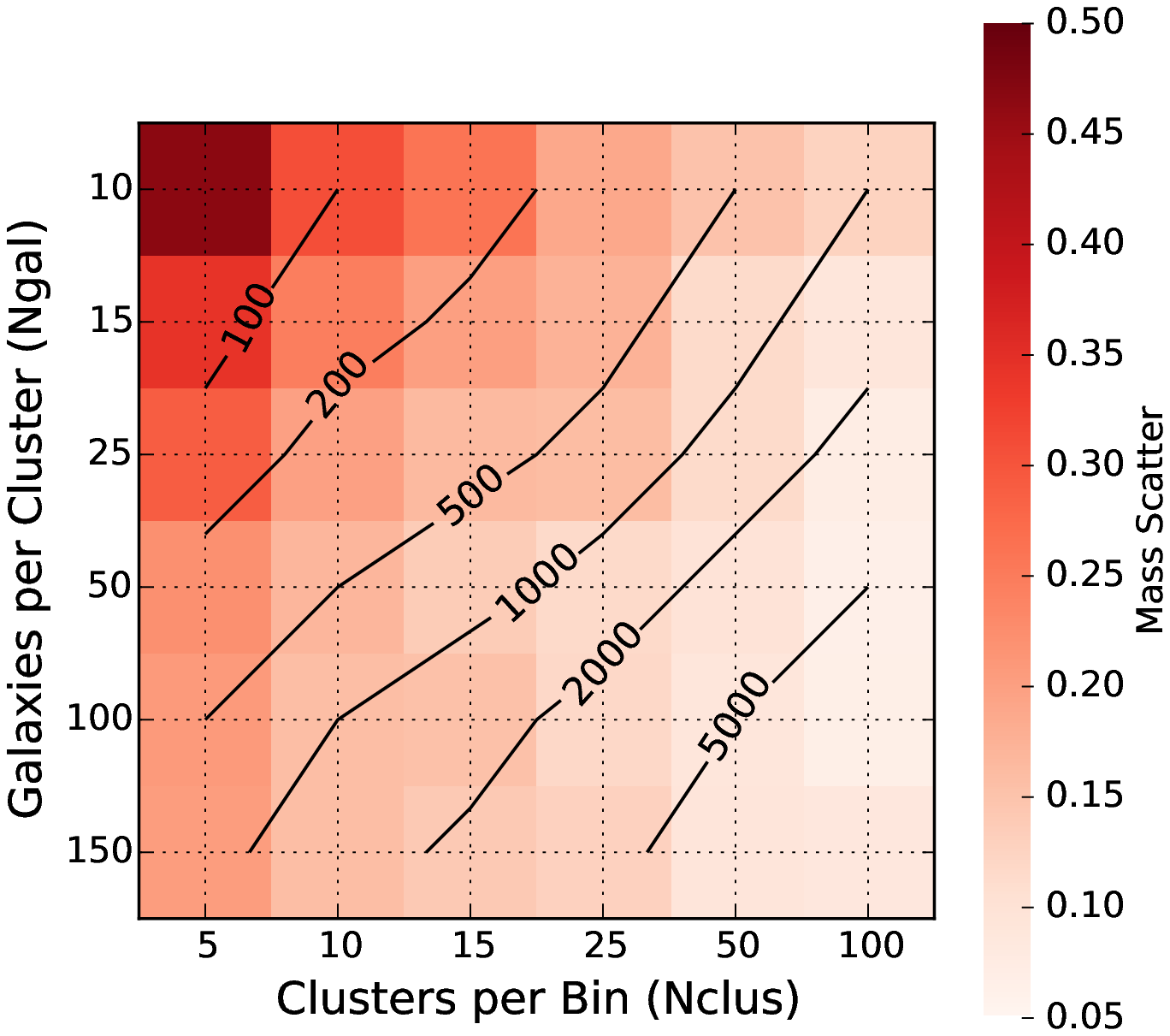}
\caption{Left: We show caustic mass bias for mass-stacked ensembles as a function of the number of random lines-of-sight to each cluster and the number of galaxies sampled from each line-of-sight. The labeled diagonal lines are contours of constant stacked phase-space richness $N_{ens}$. Right: The caustic mass scatter for mass-stacked ensembles as a function of the number of random lines-of-sight to each cluster and the number of galaxies sampled from each line-of-sight. \label{fig:BS_Bias_Scatter}}
\end{figure*}

\begin{figure}
\plotone{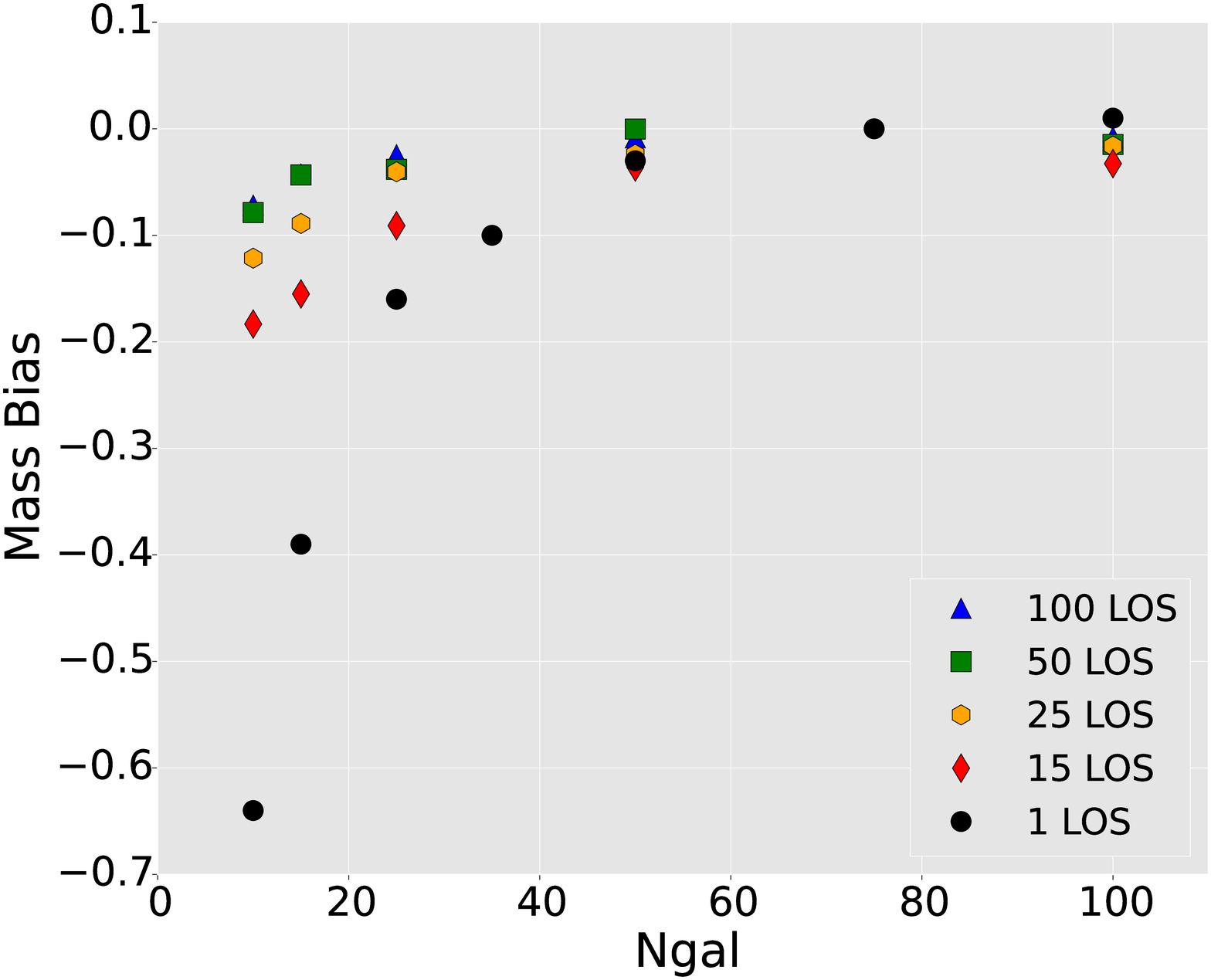}
\caption{The bias between the measured stacked mass and the average cluster mass per stacked ensemble. The different points show the bias as a function of phase-space sampling. We compare with the result from \citet{Gifford13b} that measured the average caustic mass bias for individual systems (equivalent to 1 L.O.S) as a function of phase-space sampling. Stacking works to remove the sampling bias which cannot be achieved by averaging. \label{fig:BS_Bias}}
\end{figure}

The third possible way of selecting galaxies is to assure that no galaxy appears in more than one line-of-sight phase space in the self-stacking analysis. While realistic in mimicking stacking unique systems, it results in individual line-of-sight (pre-stacked) phase-spaces that are populated with nearly all faint galaxies. While no artificial structure exists in the stacked phase-space, we do not pursue this method. We decide to use the second method to select galaxies due to its lack of artificial sub-structure with a $N_{gal}/N_{bright}$ fraction of $1/10$.

\subsubsection{Self-stacking Results}
The results of self-stacking represent an ideal scenario where the expectation mass equals the mass of the cluster being self-stacked. It is now possible to test whether the act of stacking multiple phase-space projections results in any biases. As we point out in section \ref{sec:caustic_tech}, the stacked systems are close to spherically symmetric, and any observed biases are not due to cluster shape.

We test the self-stacking performance by varying either the number of lines-of-sight $N_{los}$ stacked in phase-space or the number of bright galaxies $N_{gal}$ we sample from each line-of-sight. For example, to achieve a similar stacked richness of $N_{ens} = 500$, we can either stack $N_{los} = 10$ each with $N_{gal} = 50$ or $N_{los} = 50$ each with $N_{gal} = 10$. In Figure \ref{fig:SS_Bias_Scatter}, we vary $N_{gal}$ along the vertical direction and the $N_{los}$ in the horizontal direction. The color map represents the degree of bias (left) and scatter (right) in the recovered masses as compared with $M_{200}$ in the simulation. Contours of constant $N_{ens}$ are displayed and labeled as diagonal lines in the figure.

Of primary concern is the average degree of self-stacked mass bias relative to the mass of each halo as a function of $N_{los}$ and $N_{gal}$. We find the measured bias depends almost purely on $N_{ens}$ and asymptotically approaches the input cluster mass when $N_{ens}$ is large (lower right of Figure \ref{fig:SS_Bias_Scatter} left). On average, the stacks are unbiased to within $5\%$ of the input cluster mass when $N_{ens} > 1000$. The input cluster mass is defined as the sum of the masses of all dark matter particles in 3D space within $R_{200}$ for a given cluster. This holds even when the sampling per cluster is low ($\sim$ 10-15 galaxies) and we expect a significant low bias on individual systems \citep{Gifford13a}. We draw the conclusion that stacking multiple-phase spaces with individually unique velocity dispersions does not bias the mass inferred by the caustic technique.

Figure \ref{fig:SS_Bias_Scatter} right shows how the self-stacked mass scatter depends on $N_{ens}$. When $N_{gal} = 100$, the scatter decreases from $21 - 14\%$ as we increase the number of lines-of-sight included in each stack. This is approximately a factor of 2 less than the individual cluster mass scatter seen in \citet{Gifford13b}. We conclude that the scatter in stacked mass primarily depends on the total stacked richness which increases along an upper-left to lower-right diagonal in the parameter space.

\subsection{Mass Stacking} \label{sec:massbinstack}

In the previous section, we show that stacking projected phase-spaces from clusters of exactly the same mass does not result in a biased measurement of mass when the caustic edge is identified directly. However, this assumes the bins chosen to stack within are of infinitely small width in mass which is unphysical when stacking different clusters in simulations or observations. In an ideal scenario, we would bin clusters on an observable that has negligible scatter with mass. This would be functionally equivalent to binning on mass itself. Analytically, the expectation mass of each bin would then be
\begin{equation} \label{eq:avg_m_mass}
\langle M \rangle = \frac{\int M dn/dM \  dV/dz \ \psi(M) dM}{\int dn/dM \ dV/dz \ \psi(M) dM},
\end{equation} 
where $dn/dM$ is the halo mass function, and $\psi(M) = 1$ when $M \in [M^{min}_{200},M^{max}_{200}]$ acting as a window function for each mass bin.

We define our bins by sorting our clusters in mass, and require that each mass bin contain an equal number of clusters ($N_{clus}$). We can then vary $N_{clus}$ to study the effects of including more clusters in a stack. This is formally equivalent to varying $N_{los}$ in the self-stacking test described in section \ref{sec:self_stacking}. Following this procedure, the width of our mass bins are related to our simulated cluster sample size and is not motivated by any observational or physical reason. As a consequence of keeping $N_{clus}$ constant across all mass bins, the width of each mass bin will not be constant and adapt to the mass function of our sample. Because we have many more low than high mass clusters, the low mass bins will be far narrower than the high mass bins. The simulated cluster masses ($M_{200}$) range from $7.2\times 10^{13}$ - $2.1\times 10^{15}$.

When stacking different clusters in a mass bin, we can select the top $N_{gal}$ brightest galaxies from each cluster that are projected within $r_{200}$ and are $< \pm 4000$km/s away from the cluster. This is similar to spectroscopic follow-up in practice where bright galaxies are usually observed with higher priority given a magnitude limited cluster survey. In observations, the number of bright galaxies will vary depending on both cluster size and redshift, but in this analysis we are able to set $N_{gal}$ and test how mass scatter and bias depend on this trait. 

Figure \ref{fig:BS_Bias_Scatter} is similar to Figure \ref{fig:SS_Bias_Scatter} which shows the degree of bias (left) and scatter (right) for the mass stacking technique as compared with the average cluster mass in each ensemble. We measure these properties as a function of both $N_{gal}$ and $N_{clus}$ and find the trend follows total ensemble sampling ($N_{ens}$) that is nearly identical to the self-stacked experiment.  We find that the stacked mass estimates become unbiased once $N_{ens} > 2000$, by which is in agreement with the self-stacking results.

Seen another way, Figure \ref{fig:BS_Bias} shows the mass bias as a function of each ensemble's individual cluster sampling $N_{gal}$ compared with results from \citet{Gifford13b}. Binning on 15, 25, 50, and 100 clusters (LOS) are shown as red diamonds, orange hexagons, green squares, and blue triangles respectively. The black circles represent the results from \citet{Gifford13b} who used 100 clusters in the same Millennium simulation and measured the average bias for individual systems in using the \citet{Guo11} semi-analytic galaxy catalog. We find that stacking works to remove the bias observed in individual systems. Even when we only sample the brightest 15 galaxies per cluster, stacking 50 or more of these phase-spaces and measuring an ensemble caustic mass recovers the average mass to within 5\% as compared with the original bias seen in \citet{Gifford13b} of $-65\%$. The effects of low sampling can be seen when fewer clusters are stacked with $N_{gal} < 25$ as the method fails to identify the escape velocity edge. 

It is important to emphasize that the reduction in statistical bias for the poorly sampled ensemble clusters cannot be replicated by averaging alone. ``Ensembles'' based on averaging individual cluster caustic masses will reduce scatter, but will fail to remove the known sampling biases. Stacked ensembles produce the high phase-space sampling required for accurate caustic masses and can work even when the individual per cluster sampling is low. 

In \S \ref{sec:esc_method}, we mention that the stacked estimate of the velocity dispersion is expected to be biased relative to the average of the true cluster velocity dispersions in each stack due to mixing Gaussians. We test how these velocity dispersion biases translate to mass biases when using the stacked velocity dispersion to calibrate the stacked caustic surface. In our example in \S \ref{sec:esc_method}, we found average velocity dispersion biases of 3-10\% depending on the interloper fraction of the stacked ensemble. We find this translates to an average stacked mass bias of $\sim 3\%$ when $N_{gal} = 15$ and $N_{clus} = 50$ to directly compare with our velocity dispersion example. If we increase the sampling per cluster to $N_{gal} = 50$, the stacked mass bias increases slightly to $\sim 6\%$. This confirms our hypothesis that using a biased velocity dispersion to calibrate our stacked caustic surface will return biased stacked masses compared with the average mass in each stack. We note that this level of uncertainty is similar to our uncertainty in $\mathcal{F}_{\beta}$ (see \S \ref{sec:calib1}.)

In terms of precision, we find that the scatter in $\langle \ln M_{caustic} | M_{200}\rangle$ for ensemble clusters, decreases to 10\% for ensemble clusters with only 15 galaxies per cluster and 100 clusters per ensemble or when $N_{ens} > 2000$. Compare this to the results for individual systems, where the mass scatter is is 70\% when only 15 galaxies are used to measure the caustic mass \cite{Gifford13a}. For well sampled clusters and by using large cluster samples, one can achieve high precision mass estimates for ensemble clusters using the caustic technique (i.e., statistical uncertainties less than 5\%).

\subsection{Observable Stacking} \label{sec:observablestack}

\begin{figure*}
\plottwo{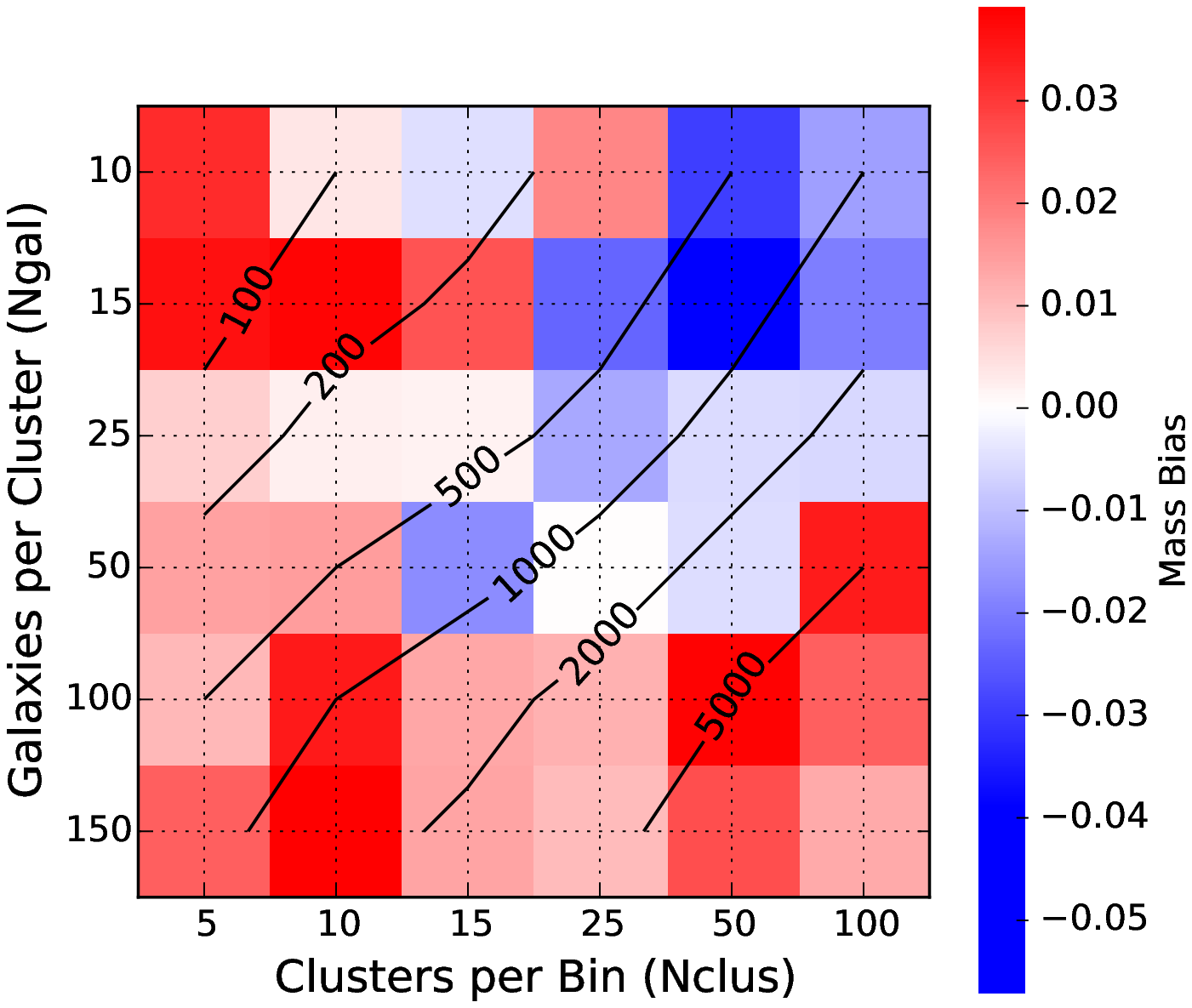}{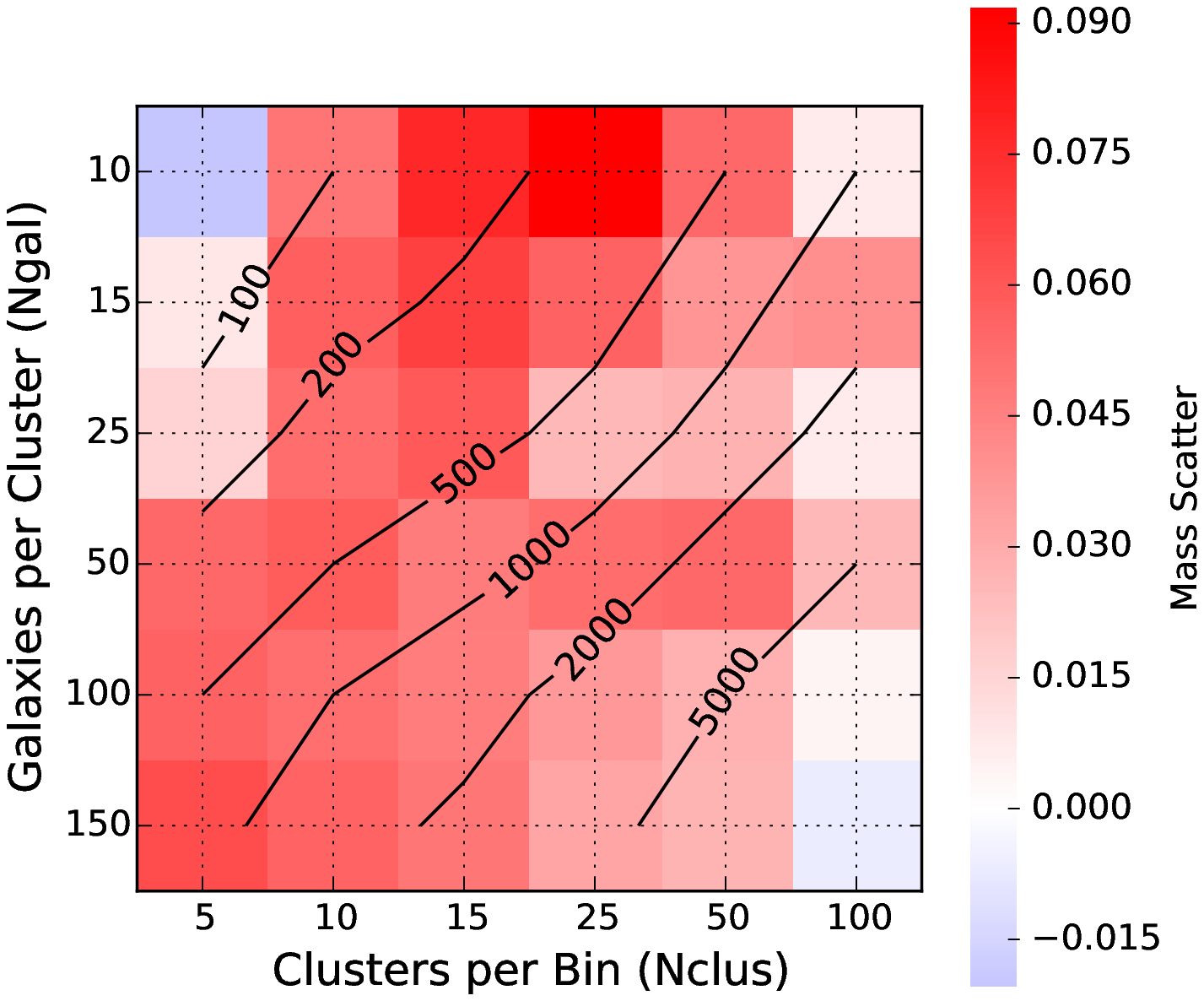}
\caption{For each bin used in Figure \ref{fig:BS_Bias_Scatter}, we calculate the scatter and the bias using perfect knowledge of the cluster masses when defining the bin membership and also inmperfect knowledge by utilizing richness as the mass-proxy. Left: The difference between the proxy-stacked and mass-stacked mass bias as a function of $N_{gal}$ and $N_{clus}$. The absolute difference is $<5\%$ when $N_{ens} > 1000$ Right: The difference between proxy-stacked and mass-stacked mass scatter as a function of $N_{gal}$ per cluster and $N_{clus}$ per stack. On average, the scatter increases by $4\%$ when binning on richness. \label{fig:mm_Bias_Scatter}}
\end{figure*}

\begin{figure*}
\centering
\plottwo{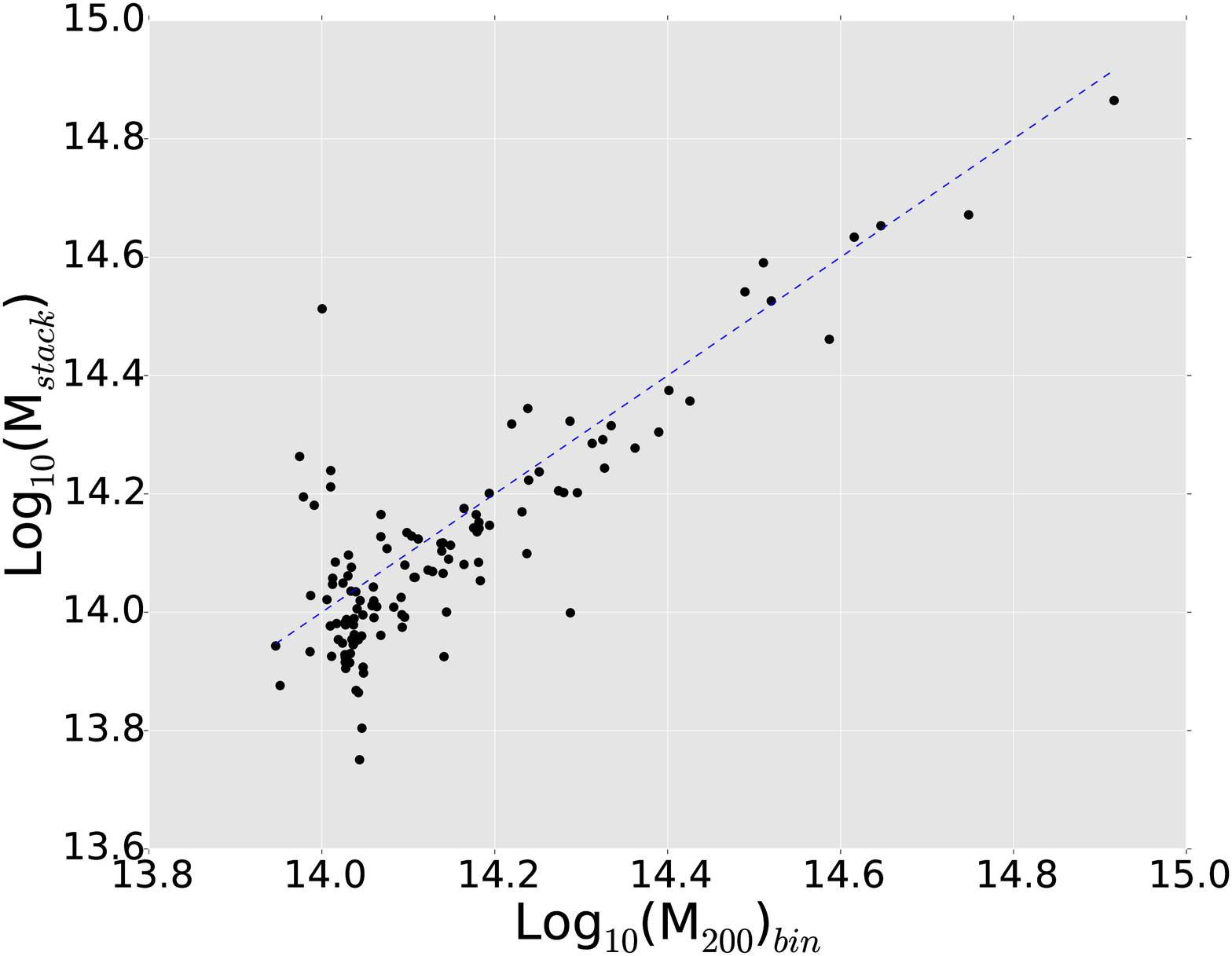}{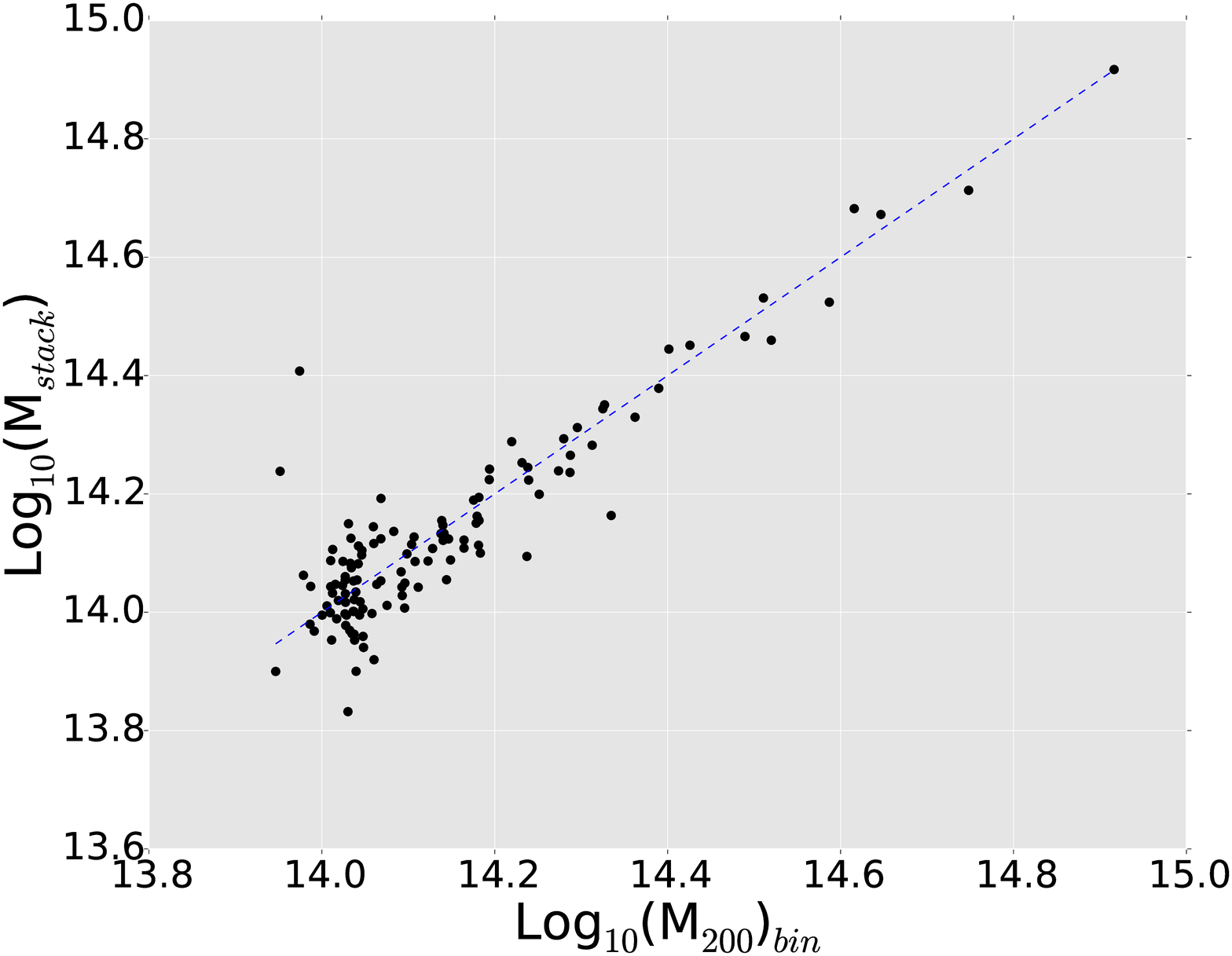}
\caption{A comparison of the average halo mass in each observable bin (Log$_{10}$ (M$_{200}$)$_{bin}$) to the respective stacked mass estimates using the caustic technique with direct surface detection. Each dot is an ensemble cluster and is constructed by binning on our richness observable. We use 50 clusters to build each ensemble and draw  $N_{gal}$ = 15 per cluster (left) and $N_{gal}$ = 50 per cluster (right). The dotted line is the one-to-one line. Note that the low richness ensembles are biased $\sim 5\%$ low, consistent with Figure \ref{fig:mm_Bias_Scatter}. }
\label{fig:onetoone}
\end{figure*}

In observations of real clusters, we must bin on cluster properties that correlate with mass. For optical surveys, examples of observables that act as mass proxies may include velocity dispersion, richness, or total luminosity. These observables often scale with mass through power law relationships that also include a degree of scatter. The relationship can be calibrated either in simulations or self-consistently in real observations, however, doing so requires an understanding of the scatter and uncertainty in the mass proxy measurement.

Scatter in mass-observable scaling relations are most commonly due to intrinsic variability and line-of-sight projection  effects like non-spherical symmetry or contamination from interlopers and large scale structure. An example of intrinsic variability is two clusters of identical mass containing different richnesses of bright galaxies within their virial radii. This is a statistical scatter that will dominate if projection effects are minimal and represents an upper limit to the relationship's precision. However, scatter between mass and observables are mostly limited by line-of-sight effects. Velocity dispersion is an excellent observable proxy of mass. In 3D, the velocity dispersion at fixed mass exhibits a minimal scatter of 5\% and is independent of cosmology \citep{Evrard08}. In projection, the scatter with mass can be as low as $15\%$ for highly sampled systems or as high as $40\%$ for systems with only a handful of spectroscopic members \citep{Gifford13b,Saro13}.

Scatter negatively affects our attempt to bin clusters based on these observable properties with the goal of closely binning on mass. Clusters with a given mass are randomly scattered between observable bins as a result of scatter in the mass-observable relationship. Because the mass function falls off sharply with mass, the up-scatter can particularly impact the high-mass bins by artificially increasing the number of clusters in those bins \citep{Lima05}. This contamination between mass bins may affect the caustic mass estimates for each ensemble and bias the resulting stacked mass relative to the true observable-mass relation expectation. As described in \citet{Rozo10}, the average mass in a richness bin is modeled as
\begin{equation} \label{eq:avg_m_proxy}
\langle M \rangle = \frac{\int M dn/dM \ dV/dz \ \langle \psi | M \rangle dM}{\int dn/dM \ dV/dz \ \langle \psi | M \rangle dM},
\end{equation}
where $dn/dM$ is the halo mass function and
\begin{equation}
\langle \psi | M \rangle = \int P(N_{200} | M) \psi(N_{200}) dN_{200}.
\end{equation}
Here, $\psi(N_{200}) = 1$ when $N_{200} \in [N^{min}_{200},N^{max}_{200}]$ per bin, and $P(N_{200} | M)$ is the probability a cluster with mass $M$ has a richness $N_{200}$. $\langle \psi | M \rangle$ then equals the probability that a cluster of mass $M$ is observed within the richness bin controlled by $\psi$.

Observable-mass relations with extremely tight scatter that limit the effects listed above do exist as tested in high-resolution Nbody and gas dynamics simulations. \citet{Kravtsov06} found that the product of the gas mass and temperature observables ($Y_X$) show a remarkably low scatter of $<10\%$ with $M_{500}$. Unfortunately, these observables are often not available for the majority of optically identified clusters. Instead, we focus on optical observables such as cluster richness which is known to correlate tightly with mass. \citet{Rozo09} used SDSS clusters to estimate the scatter on their matched-filter richness estimator ($\lambda$) and found $<25\%$ log-scatter with $M_{200}$ for clusters with $N = 40$. In our simulations, we utilize a simple richness estimator for a background subtracted galaxy count within a projected aperture on the sky equal to $r_{200}$. We estimate the background counts by sampling an area of sky 5x the cluster aperture, normalize by area, and take the difference from the cluster counts to measure $N_{200}$. Using the true values of $M_{200}$ in the simulation and comparing with our estimated values of $N_{200}$, we measure a log-scatter in $P(N_{200} | M)$ equal to 20\%, which is realistic when compared to \citep{Rozo09}. This level of scatter in the mass-observable relation is also realistic for future spectroscopic surveys which measure spectroscopic members for tens of galaxies per cluster (see Section \ref{sec:Discussion}).

We apply our estimates of $N_{200}$ to our stacking framework in a similar systematic fashion as \S \ref{sec:massbinstack}. The two variables of interest are the number of galaxies we sample from each cluster ($N_{gal}$)and the number of clusters we stack per bin ($N_{clus}$). However, instead of building our stacks based on binning according to the true underlying cluster mass, we do so based on our optical richness observable. Note that we do not use or require the use of the underlying mass-observable relation in the analysis. We measure the richnesses in a realistic fashion and build the ensembles directly on the observed richnesses. The comparison is then made between the ensemble caustic mass and the average of the true underlying halo masses for those same systems in each richness bin.

In Figure \ref{fig:mm_Bias_Scatter} (left) we present how the sampling and mass-scatter affect the stacked caustic mass bias for different types of sampling and binning relative to the results in Figure \ref{fig:BS_Bias_Scatter}. We observe that, when binning on an observable with realistic scatter, the stacked caustic mass recovers the average mass of the stack with the same accuracy as our perfectly binned sample to within $\pm 3\%$. This result indicates that modest scatter in the mass-observable relationship does not impact the stacked caustic technique's ability to recover average masses in each observable bin. The scatter between the ensemble mass estimates and the average mass per bin is also minimally affected. In Figure \ref{fig:mm_Bias_Scatter} (right), we quantify the increase in scatter due to the increase in mass mixing in each observable bin. Across the range of galaxy and cluster sampling per bin, we see an increase in the scatter by $\sim$ 5\% when there exists scatter between the true mass and observational mass-proxy used to bin. Overall, the minimum mass scatter reached for observable stacked ensembles is $< 10\%$ in highly sampled clusters using large cluster samples. 

In Figure \ref{fig:onetoone}, we show one-to-one plots of average cluster mass vs the stacked mass per bin with $N_{clus} = 50$ in both panels. The left panel stacks are built by sampling $N_{gal} = 15$ per cluster, and the right panel stacks by sampling $N_{gal} = 50$ per cluster.  As shown in \citep{Gifford13a}, Figure \ref{fig:onetoone} should show a 40\% negative bias if one used the individual cluster masses and averaged over the large sample of clusters. The stacked bias is smaller than 5\%. 

On the other hand it is possible for stacked ensemble estimates to fail and return erroneous values. These are the outliers in \ref{fig:onetoone}. However, even when sampling only 15 galaxies per cluster, it happens rarely. With a modest degree of mass-mixing due to observable-mass scatter, only a few percent of our stacked estimates are classified as ``failed stacks". These outliers are in the low mass end where our interloper rejection fails. Based on the low level of bias and scatter, combined with the lack of outliers in the stacked estimates, we can conclude that the stacking technique is robust to binning on a mass-proxy such as richness. 

\subsection{Cosmological and Astrophysical Dependencies}

Throughout this work, we have utilized a single simulated galaxy catalog based on a single underlying cosmological simulation \citep{Guo11,Henriques12}. It is worth discussing how other mock galaxy prescriptions and other cosmologies might affect our results. 

Previous authors have studied the issues of dynamical mass tracers and cosmology. For instance, \citet{Evrard08} showed that the 3D virial-to-mass scaling relation is independent of cosmology, based a comparison against a number of different cosmologies. The same can be said for the standard caustic technique, which is governed by the Poisson equation and calibrated by the virial equation (Diaferio 1999 and Section 
\ref{sec:caustic_tech}). However, all of our measurements are made on projected data where the dominant concern for the surface calibration is the normalization of the power spectrum (i.e., $\sigma_8$, which sets the normalization on 8Mpc scales). Projected interlopers are the likely cause for additional systematic biases in the stacked surface measurement. For example, there are $\sim 30\%$ more clusters above $M_{200} = 5\times10^{13}$M$_{\odot}$ when $\sigma_8 = 0.9$ compared to $\sigma_8 =0.8$. The additional clusters in a high $\sigma_8$ Universe could result in contaminated phase spaces, which in turn could change the percentile used to identify the stacked caustic surface.  

As a final test, we conduct our entire analysis using the MICE mock sky catalogs \citep{Fosalba2008, Crocce2010}. Like the Millennium data, the MICE data stems from a dark matter N-body particle simulation using GADGET-2. The cosmology used in the MICE simulation is almost identical to that used in the Millennium simulation ($\Omega_{m}= 0.25, \Omega_{\Lambda} = 0.75, \Omega_{baryon} = 0.045$ for Millennium and $\Omega_{baryon} = 0.044$ for MICE). 

There are two key differences between the MICE galaxy catalogs and the Millennium galaxy catalogs. First, the normalization of the power spectrum in the MICE simulations is $\sigma_8 = 0.8$ while for the Millennium it is $\sigma_8 = 0.9$. Second, while the Henriques et al. (2012) light-cone is developed using a semi-analytic representation of galaxy formation and evolution, the MICE light-cone mock galaxy catalogs are developed using a hybrid Halo Occupation Distribution and Halo Abundance Matching prescription to populate Friends-of-Friends dark matter halos. The details of how the galaxies are assigned to halos and how their magnitudes are prescribed in the MICE version 1 simulations is discussed in \cite{Carretero2015}.

We then conduct the identical exercise on the MICE-based data that we used to build the ensemble caustics as discussed in Section \ref{sec:massbinstack} for the Millennium-based data. We use the surface calibration as determined from the Millennium and presented in Section \ref{sec:esc_method}. Note that there are significantly fewer massive halos in the MICE catalog, due to the smaller areal coverage of only 1/8$^{th}$ the total sky and the smaller power spectrum normalization. However, for most of the bins defined in Figure \ref{fig:BS_Bias_Scatter}, we can make a direct comparison.

When comparing the stacking results from the MICE to the Millennium simulations, we find that the mean absolute deviation of the measured ensemble scatter between the Millennium and the MICE samples is $0.014 \pm{0.003}$. The mean absolute deviation of the measured ensemble bias between the Millennium and the MICE samples is $-0.023\pm{0.005}$. These small differences in the measured bias and scatter between the two simulations establish the baseline of the robustness of the caustic surface definition technique against large variations in astrophysical and cosmological variations. A more detailed study would require a significant number of light-cone mock galaxy catalogs constructed against a wider variation in the cosmological parameters. Such data does not yet exist.

\section{Discussion and Conclusions}
\label{sec:Discussion}
Measuring dynamical cluster masses when $N_{gal} < 25$ can result in statistical and systematic biases that make it impossible to self-calibrate mass-observable scaling relations through averaging \citep{White10,Wu13,Gifford13b}. Stacking works to eliminate statistical biases present in individual cluster observations by building an ensemble system with sufficient sampling to measure average properties of the stacked system.  However, using stacking methods to infer the average properties of clusters can be a messy process that depends on prior decisions of how to bin, which observable to bin on, the cluster finder and survey used, etc.

We have used halos in an all-sky light cone built on the Millennium Simulation \citep{Guo11,Henriques12} to define a complete and pure, low redshift ($z \le 0.15$) cluster sample. We then develop and characterize a caustic stacking technique to infer ensemble cluster masses. We find that the surface calibration in the caustic technique needs to be modified to avoid velocity biases. The standard virial-based surface calibration imparts a 5-10\% statistical mass bias into the ensemble masses, which is similar to the the uncertainty on the caustic calibration term, $\mathcal{F}_{\beta}$. To solve this problem, we develop and test a new calibration of the caustic surface which leverages the well-populated ensemble phase-spaces. 

We evaluate our new stacking procedure in terms of bias and scatter for three scenarios: (1) an ideal situation where the mass of an ensemble cluster is perfectly known and defined by a single cluster (self-stacking); (2) when the mass of the clusters are known but the ensemble contains a distribution of cluster masses; (3) when the mass of the clusters is inferred through a proxy with scatter and the ensemble is defined based on this proxy. This study should be viewed as a baseline ability of the stacked caustic technique to recover average cluster masses. The completeness, purity, and mis-centering of cluster finding methods may well influence the results presented here \citep{Miller05,Rozo09}.

When creating the ensemble clusters, we vary the number of galaxies sampled per cluster and the number of clusters per bin. We find that the agreement between the stacked caustic estimate and the average true mass of the bin depends on the total sampling in the stacked cluster projected phase-space. Once we achieve a level of sampling of $\sim 1000$ tracers within the phase-space, our estimates are unbiased to within $5\%$. The bias follows the iso-sampling contours very closely, implying that achieving precise stacked estimates depends solely on the total number of galaxies in the stacked phase-space.  Our tests also reveal that stacking results in very low mass scatter relative to the average cluster mass per bin: $\lesssim 10\%$ when our phase-space sampling is high ($N_{ens} > 5000$).

We incorporate mass scatter into the analysis through an observable mass proxy via cluster richnesses to demonstrate the effects of adding uncertainty to the binning process. When binning on richness, we are imposing a log mass scatter of 20\% (intrinsically built into the Henriques et al. 2012 mock galaxy catalogs). We find that the ensemble stacked caustics cluster masses can recover the underlying binned mass averages to within $\sim 5\%$, matching the scenario where we assume no scatter between observable and mass. This accuracy is independent of cluster mass regardless of binning procedure. We find that the log scatter in the recovered masses increases by 5-10\% (see Figure \ref{fig:mm_Bias_Scatter}).

Cosmological constraints based on stacking observed data within bins based on the observable do not yet fully utilize the data. As an example, mass scatter plays an important role in the observed abundance function \citep{Rozo10}. By using the stacking techniques described here, in combination with individual caustic cluster masses, one could directly characterize the scatter and bias in mass using the data alone. In this paper, we show that caustic masses of ensemble clusters in projected data are both accurate and precise (i.e., within 5\% in $\langle \ln M_{caustic} | M_{200}\rangle$). This high level of accuracy and precision provides the means to measure the sample-wide statistical bias and scatter by directly using the individual cluster measurements and comparing to the stacked measurements. 

Our efforts in this paper coincide with current surveys like the Dark Energy Survey \citep{FlaugherDECam2015}, the Baryon Acoustic Oscillation Survey and its extension eBOSS (extendedBOSS is part of a program of post-2014 surveys on the Sloan telescope), and the Dark Energy Spectroscopic Instrument \citep{Levi13}. Each of these surveys is providing (or will provide) a wealth of new photometric and spectroscopic data for clusters over a wide range of mass and redshift. These new data and the stacking techniques described here will enable precision stacked dynamical cluster mass estimates to redshifts $z \sim 0.7$ for cosmological analyses \citep{Vikhlinin09}.

\section{Acknowledgements}
The authors made use of the FLUX High Performance Computing Cluster at the University of Michigan. The Millennium Simulation databases used in this paper and the web application providing online access to them were constructed as part of the activities of the German Astrophysical Virtual Observatory (GAVO). The authors want to thank Alejo Stark, Jessica Kellar, and August Evrard for their helpful comments and discussion. This material is based upon work supported by the National Science Foundation under Grant No. 1311820 and Grant No. 1256260.

\bibliographystyle{apj}

\begin{thebibliography}{52}
\expandafter\ifx\csname natexlab\endcsname\relax\def\natexlab#1{#1}\fi

\bibitem[{{Alpaslan} {et~al.}(2012){Alpaslan}, {Robotham}, {Driver}, {Norberg},
  {Peacock}, {Baldry}, {Bland-Hawthorn}, {Brough}, {Hopkins}, {Kelvin},
  {Liske}, {Loveday}, {Merson}, {Nichol}, \& {Pimbblet}}]{Alpaslan12}
{Alpaslan}, M., {Robotham}, A.~S.~G., {Driver}, S., {et~al.} 2012, \mnras, 426,
  2832

\bibitem[{{Andreon} \& {Hurn}(2010)}]{Andreon10}
{Andreon}, S., \& {Hurn}, M.~A. 2010, \mnras, 404, 1922

\bibitem[{{Becker} \& {Kravtsov}(2011)}]{Becker11}
{Becker}, M.~R., \& {Kravtsov}, A.~V. 2011, \apj, 740, 25

\bibitem[{{Becker} {et~al.}(2007){Becker}, {McKay}, {Koester}, {Wechsler},
  {Rozo}, {Evrard}, {Johnston}, {Sheldon}, {Annis}, {Lau}, {Nichol}, \&
  {Miller}}]{Becker07}
{Becker}, M.~R., {McKay}, T.~A., {Koester}, B., {et~al.} 2007, \apj, 669, 905

\bibitem[{{Biesiadzinski} {et~al.}(2012){Biesiadzinski}, {McMahon}, {Miller},
  {Nord}, \& {Shaw}}]{Biesiadzinski12}
{Biesiadzinski}, T., {McMahon}, J., {Miller}, C.~J., {Nord}, B., \& {Shaw}, L.
  2012, \apj, 757, 1

\bibitem[{{Binney} \& {Tremaine}(1987)}]{Binney87}
{Binney}, J., \& {Tremaine}, S. 1987, {Galactic dynamics}

\bibitem[{{Biviano} \& {Girardi}(2003)}]{Biviano03}
{Biviano}, A., \& {Girardi}, M. 2003, \apj, 585, 205

\bibitem[{{Biviano} {et~al.}(2006){Biviano}, {Murante}, {Borgani}, {Diaferio},
  {Dolag}, \& {Girardi}}]{Biviano06}
{Biviano}, A., {Murante}, G., {Borgani}, S., {et~al.} 2006, \aap, 456, 23

\bibitem[{{Carlberg} {et~al.}(1997){Carlberg}, {Yee}, {Ellingson}, {Morris},
  {Abraham}, {Gravel}, {Pritchet}, {Smecker-Hane}, {Hartwick}, {Hesser},
  {Hutchings}, \& {Oke}}]{Carlberg97}
{Carlberg}, R.~G., {Yee}, H.~K.~C., {Ellingson}, E., {et~al.} 1997, \apjl, 485,
  L13

\bibitem[{{Carretero} {et~al.}(2015){Carretero}, {Castander}, {Gazta{\~n}aga},
  {Crocce}, \& {Fosalba}}]{Carretero2015}
{Carretero}, J., {Castander}, F.~J., {Gazta{\~n}aga}, E., {Crocce}, M., \&
  {Fosalba}, P. 2015, \mnras, 447, 646

\bibitem[{{Crocce} {et~al.}(2010){Crocce}, {Fosalba}, {Castander}, \&
  {Gazta{\~n}aga}}]{Crocce2010}
{Crocce}, M., {Fosalba}, P., {Castander}, F.~J., \& {Gazta{\~n}aga}, E. 2010,
  \mnras, 403, 1353

\bibitem[{{Dehnen}(1993)}]{Dehnen93}
{Dehnen}, W. 1993, \mnras, 265, 250

\bibitem[{{Diaferio}(1999)}]{Diaferio99}
{Diaferio}, A. 1999, \mnras, 309, 610

\bibitem[{{Diaferio} \& {Geller}(1997)}]{Diaferio97}
{Diaferio}, A., \& {Geller}, M.~J. 1997, \apj, 481, 633

\bibitem[{{Dietrich} {et~al.}(2014){Dietrich}, {Zhang}, {Song}, {Davis},
  {McKay}, {Baruah}, {Becker}, {Benoist}, {Busha}, {da Costa}, {Hao}, {Maia},
  {Miller}, {Ogando}, {Romer}, {Rozo}, {Rykoff}, \& {Wechsler}}]{Dietrich14}
{Dietrich}, J.~P., {Zhang}, Y., {Song}, J., {et~al.} 2014, \mnras, 443, 1713

\bibitem[{{Einasto}(1969)}]{Einasto69}
{Einasto}, J. 1969, Astronomische Nachrichten, 291, 97

\bibitem[{{Evrard} {et~al.}(2008){Evrard}, {Bialek}, {Busha}, {White}, {Habib},
  {Heitmann}, {Warren}, {Rasia}, {Tormen}, {Moscardini}, {Power}, {Jenkins},
  {Gao}, {Frenk}, {Springel}, {White}, \& {Diemand}}]{Evrard08}
{Evrard}, A.~E., {Bialek}, J., {Busha}, M., {et~al.} 2008, \apj, 672, 122

\bibitem[{{Flaugher} {et~al.}(2015){Flaugher}, {Diehl}, {Honscheid}, {Abbott},
  {Alvarez}, {Angstadt}, {Annis}, {Antonik}, {Ballester}, {Beaufore},
  {Bernstein}, {Bernstein}, {Bigelow}, {Bonati}, {Boprie}, {Brooks},
  {Buckley-Geer}, {Campa}, {Cardiel-Sas}, {Castander}, {Castilla}, {Cease},
  {Cela-Ruiz}, {Chappa}, {Chi}, {Cooper}, {da Costa}, {Dede}, {Derylo},
  {DePoy}, {de Vicente}, {Doel}, {Drlica-Wagner}, {Eiting}, {Elliott}, {Emes},
  {Estrada}, {Fausti Neto}, {Finley}, {Flores}, {Frieman}, {Gerdes},
  {Gladders}, {Gregory}, {Gutierrez}, {Hao}, {Holland}, {Holm}, {Huffman},
  {Jackson}, {James}, {Jonas}, {Karcher}, {Karliner}, {Kent}, {Kessler},
  {Kozlovsky}, {Kron}, {Kubik}, {Kuehn}, {Kuhlmann}, {Kuk}, {Lahav}, {Lathrop},
  {Lee}, {Levi}, {Lewis}, {Li}, {Mandrichenko}, {Marshall}, {Martinez},
  {Merritt}, {Miquel}, {Munoz}, {Neilsen}, {Nichol}, {Nord}, {Ogando}, {Olsen},
  {Palio}, {Patton}, {Peoples}, {Plazas}, {Rauch}, {Reil}, {Rheault}, {Roe},
  {Rogers}, {Roodman}, {Sanchez}, {Scarpine}, {Schindler}, {Schmidt},
  {Schmitt}, {Schubnell}, {Schultz}, {Schurter}, {Scott}, {Serrano}, {Shaw},
  {Smith}, {Soares-Santos}, {Stefanik}, {Stuermer}, {Suchyta}, {Sypniewski},
  {Tarle}, {Thaler}, {Tighe}, {Tran}, {Tucker}, {Walker}, {Wang}, {Watson},
  {Weaverdyck}, {Wester}, {Woods}, \& {Yanny}}]{FlaugherDECam2015}
{Flaugher}, B., {Diehl}, H.~T., {Honscheid}, K., {et~al.} 2015, ArXiv e-prints

\bibitem[{{Fosalba} {et~al.}(2008){Fosalba}, {Gazta{\~n}aga}, {Castander}, \&
  {Manera}}]{Fosalba2008}
{Fosalba}, P., {Gazta{\~n}aga}, E., {Castander}, F.~J., \& {Manera}, M. 2008,
  \mnras, 391, 435

\bibitem[{{Geller} {et~al.}(2013){Geller}, {Diaferio}, {Rines}, \&
  {Serra}}]{Geller13}
{Geller}, M.~J., {Diaferio}, A., {Rines}, K.~J., \& {Serra}, A.~L. 2013, \apj,
  764, 58

\bibitem[{{Gifford} {et~al.}(2013){Gifford}, {Miller}, \& {Kern}}]{Gifford13b}
{Gifford}, D., {Miller}, C., \& {Kern}, N. 2013, \apj, 773, 116

\bibitem[{{Gifford} \& {Miller}(2013)}]{Gifford13a}
{Gifford}, D., \& {Miller}, C.~J. 2013, \apjl, 768, L32

\bibitem[{{Gruen} {et~al.}(2015){Gruen}, {Seitz}, {Becker}, {Friedrich}, \&
  {Mana}}]{Gruen15}
{Gruen}, D., {Seitz}, S., {Becker}, M.~R., {Friedrich}, O., \& {Mana}, A. 2015,
  \mnras, 449, 4264

\bibitem[{{Guo} {et~al.}(2011){Guo}, {White}, {Boylan-Kolchin}, {De Lucia},
  {Kauffmann}, {Lemson}, {Li}, {Springel}, \& {Weinmann}}]{Guo11}
{Guo}, Q., {White}, S., {Boylan-Kolchin}, M., {et~al.} 2011, \mnras, 413, 101

\bibitem[{{Henriques} {et~al.}(2012){Henriques}, {White}, {Lemson}, {Thomas},
  {Guo}, {Marleau}, \& {Overzier}}]{Henriques12}
{Henriques}, B.~M.~B., {White}, S.~D.~M., {Lemson}, G., {et~al.} 2012, \mnras,
  421, 2904

\bibitem[{{Hoekstra} {et~al.}(2015){Hoekstra}, {Herbonnet}, {Muzzin}, {Babul},
  {Mahdavi}, {Viola}, \& {Cacciato}}]{Hoekstra15}
{Hoekstra}, H., {Herbonnet}, R., {Muzzin}, A., {et~al.} 2015, \mnras, 449, 685

\bibitem[{{Iannuzzi} \& {Dolag}(2012)}]{Iannuzzi12}
{Iannuzzi}, F., \& {Dolag}, K. 2012, \mnras, 427, 1024

\bibitem[{{Knebe} {et~al.}(2011){Knebe}, {Knollmann}, {Muldrew}, {Pearce},
  {Aragon-Calvo}, {Ascasibar}, {Behroozi}, {Ceverino}, {Colombi}, {Diemand},
  {Dolag}, {Falck}, {Fasel}, {Gardner}, {Gottl{\"o}ber}, {Hsu}, {Iannuzzi},
  {Klypin}, {Luki{\'c}}, {Maciejewski}, {McBride}, {Neyrinck}, {Planelles},
  {Potter}, {Quilis}, {Rasera}, {Read}, {Ricker}, {Roy}, {Springel}, {Stadel},
  {Stinson}, {Sutter}, {Turchaninov}, {Tweed}, {Yepes}, \& {Zemp}}]{Knebe11}
{Knebe}, A., {Knollmann}, S.~R., {Muldrew}, S.~I., {et~al.} 2011, \mnras, 415,
  2293

\bibitem[{{Kravtsov} {et~al.}(2006){Kravtsov}, {Vikhlinin}, \&
  {Nagai}}]{Kravtsov06}
{Kravtsov}, A.~V., {Vikhlinin}, A., \& {Nagai}, D. 2006, \apj, 650, 128

\bibitem[{{Levi} {et~al.}(2013){Levi}, {Bebek}, {Beers}, {Blum}, {Cahn},
  {Eisenstein}, {Flaugher}, {Honscheid}, {Kron}, {Lahav}, {McDonald}, {Roe},
  {Schlegel}, \& {representing the DESI collaboration}}]{Levi13}
{Levi}, M., {Bebek}, C., {Beers}, T., {et~al.} 2013, ArXiv e-prints

\bibitem[{{Lima} \& {Hu}(2005)}]{Lima05}
{Lima}, M., \& {Hu}, W. 2005, \prd, 72, 043006

\bibitem[{{Meneghetti} {et~al.}(2010){Meneghetti}, {Rasia}, {Merten},
  {Bellagamba}, {Ettori}, {Mazzotta}, {Dolag}, \& {Marri}}]{Meneghetti10}
{Meneghetti}, M., {Rasia}, E., {Merten}, J., {et~al.} 2010, \aap, 514, A93

\bibitem[{{Merritt} {et~al.}(2006){Merritt}, {Graham}, {Moore}, {Diemand}, \&
  {Terzi{\'c}}}]{Merritt06}
{Merritt}, D., {Graham}, A.~W., {Moore}, B., {Diemand}, J., \& {Terzi{\'c}}, B.
  2006, \aj, 132, 2685

\bibitem[{{Merten} {et~al.}(2015)}]{Merten15}
{Merten}, J., {et~al.} 2015, \apj, 806, 4

\bibitem[{{Miller} {et~al.}(2016){Miller}, {Stark}, {Gifford}, \&
  {Kern}}]{Miller16}
{Miller}, C.~J., {Stark}, A., {Gifford}, D., \& {Kern}, N. 2016, \apj, 822, 41

\bibitem[{{Miller} {et~al.}(2005){Miller}, {Nichol}, {Reichart}, {Wechsler},
  {Evrard}, {Annis}, {McKay}, {Bahcall}, {Bernardi}, {Boehringer}, {Connolly},
  {Goto}, {Kniazev}, {Lamb}, {Postman}, {Schneider}, {Sheth}, \&
  {Voges}}]{Miller05}
{Miller}, C.~J., {Nichol}, R.~C., {Reichart}, D., {et~al.} 2005, \aj, 130, 968

\bibitem[{{Navarro} {et~al.}(1997){Navarro}, {Frenk}, \& {White}}]{NFW97}
{Navarro}, J.~F., {Frenk}, C.~S., \& {White}, S.~D.~M. 1997, \apj, 490, 493

\bibitem[{{Planck Collaboration}(2011)}]{Plank11}
{Planck Collaboration}. 2011, \aap, 536, A10

\bibitem[{{Rasia} {et~al.}(2006){Rasia}, {Ettori}, {Moscardini}, {Mazzotta},
  {Borgani}, {Dolag}, {Tormen}, {Cheng}, \& {Diaferio}}]{Rasia06}
{Rasia}, E., {Ettori}, S., {Moscardini}, L., {et~al.} 2006, \mnras, 369, 2013

\bibitem[{{Rines} {et~al.}(2003){Rines}, {Geller}, {Kurtz}, \&
  {Diaferio}}]{Rines03}
{Rines}, K., {Geller}, M.~J., {Kurtz}, M.~J., \& {Diaferio}, A. 2003, \aj, 126,
  2152

\bibitem[{{Rozo} {et~al.}(2009){Rozo}, {Rykoff}, {Evrard}, {Becker}, {McKay},
  {Wechsler}, {Koester}, {Hao}, {Hansen}, {Sheldon}, {Johnston}, {Annis}, \&
  {Frieman}}]{Rozo09}
{Rozo}, E., {Rykoff}, E.~S., {Evrard}, A., {et~al.} 2009, \apj, 699, 768

\bibitem[{{Rozo} {et~al.}(2010){Rozo}, {Wechsler}, {Rykoff}, {Annis}, {Becker},
  {Evrard}, {Frieman}, {Hansen}, {Hao}, {Johnston}, {Koester}, {McKay},
  {Sheldon}, \& {Weinberg}}]{Rozo10}
{Rozo}, E., {Wechsler}, R.~H., {Rykoff}, E.~S., {et~al.} 2010, \apj, 708, 645

\bibitem[{{Saro} {et~al.}(2013){Saro}, {Mohr}, {Bazin}, \& {Dolag}}]{Saro13}
{Saro}, A., {Mohr}, J.~J., {Bazin}, G., \& {Dolag}, K. 2013, \apj, 772, 47

\bibitem[{{Sereno} {et~al.}(2016){Sereno}, {Fedeli}, \&
  {Moscardini}}]{Sereno16}
{Sereno}, M., {Fedeli}, C., \& {Moscardini}, L. 2016, J. Cosmology Astropart.
  Phys., 1, 042

\bibitem[{{Serra} {et~al.}(2011){Serra}, {Diaferio}, {Murante}, \&
  {Borgani}}]{Serra11}
{Serra}, A.~L., {Diaferio}, A., {Murante}, G., \& {Borgani}, S. 2011, \mnras,
  412, 800

\bibitem[{{Springel} {et~al.}(2001){Springel}, {White}, {Tormen}, \&
  {Kauffmann}}]{Springel01}
{Springel}, V., {White}, S.~D.~M., {Tormen}, G., \& {Kauffmann}, G. 2001,
  \mnras, 328, 726

\bibitem[{{Springel} {et~al.}(2005){Springel}, {White}, {Jenkins}, {Frenk},
  {Yoshida}, {Gao}, {Navarro}, {Thacker}, {Croton}, {Helly}, {Peacock}, {Cole},
  {Thomas}, {Couchman}, {Evrard}, {Colberg}, \& {Pearce}}]{Springel05}
{Springel}, V., {White}, S.~D.~M., {Jenkins}, A., {et~al.} 2005, \nat, 435, 629

\bibitem[{{Stark} {et~al.}(2016){Stark}, {Miller}, \& {Gifford}}]{Stark16}
{Stark}, A., {Miller}, C.~J., \& {Gifford}, D. 2016, \apj, 830, 109

\bibitem[{{Svensmark} {et~al.}(2015){Svensmark}, {Wojtak}, \&
  {Hansen}}]{Svensmark14}
{Svensmark}, J., {Wojtak}, R., \& {Hansen}, S.~H. 2015, \mnras, 448, 1644

\bibitem[{{Vikhlinin} {et~al.}(2009){Vikhlinin}, {Kravtsov}, {Burenin},
  {Ebeling}, {Forman}, {Hornstrup}, {Jones}, {Murray}, {Nagai}, {Quintana}, \&
  {Voevodkin}}]{Vikhlinin09}
{Vikhlinin}, A., {Kravtsov}, A.~V., {Burenin}, R.~A., {et~al.} 2009, \apj, 692,
  1060

\bibitem[{{White} {et~al.}(2010){White}, {Cohn}, \& {Smit}}]{White10}
{White}, M., {Cohn}, J.~D., \& {Smit}, R. 2010, \mnras, 408, 1818

\bibitem[{{Wu} {et~al.}(2013){Wu}, {Hahn}, {Evrard}, {Wechsler}, \&
  {Dolag}}]{Wu13}
{Wu}, H.-Y., {Hahn}, O., {Evrard}, A.~E., {Wechsler}, R.~H., \& {Dolag}, K.
  2013, \mnras, 436, 460

\end{thebibliography}

\end{document}